\documentclass[twocolumn,superscriptaddress,prb,aps,floats,floatfix,showpacs,citeautoscript]{revtex4}
\usepackage{graphicx}
\usepackage{bm}
\usepackage{times}

\begin{document}
%
%
\newcommand{\be}{\begin{equation}}
\newcommand{\ee}{\end{equation}}
\newcommand{\bea}{\begin{eqnarray}}
\newcommand{\eea}{\end{eqnarray}}
\newcommand{\beann}{\begin{eqnarray*}}
\newcommand{\eeann}{\end{eqnarray*}}
\newcommand{\bma}{\begin{array}{cc}}
\newcommand{\ema}{\end{array}}
\newcommand{\fr}{\frac}
\newcommand{\ra}{\rangle}
\newcommand{\la}{\langle}
\newcommand{\li}{\left}
\newcommand{\re}{\right}
\newcommand{\ri}{\right}
\newcommand{\uarr}{\uparrow}
\newcommand{\darr}{\downarrow}
\newcommand{\df}{\stackrel{\rm def}{=}}
\newcommand{\nn}{\nonumber}
\newcommand{\dpl}{\displaystyle}
\newcommand{\p}{\partial}

\newcommand{\alp}{\alpha}
\newcommand{\sig}{\sigma}
\newcommand{\eps}{\epsilon}
\newcommand{\xsi}{\xi}
\newcommand{\lam}{\lambda}
\newcommand{\ny}{\nu}

\newcommand{\HamO}{\hat{H}_0}
\newcommand{\Ham}{\hat{H}}
\newcommand{\HamV}{\hat{V}_c}
\newcommand{\seteps}{ \{ \eps \} }
\newcommand{\setlam}{ \{ \lam \} }
\newcommand{\ef}{E_F}
\newcommand{\Deltaml}{d}
\newcommand{\Deltaov}{\Delta}
\newcommand{\Deltamubargamma}{ \Deltaov_{\bar{\mu}\gamma} }
\newcommand{\Deltaibarj}{ \Deltaov_{\bar{i} j} }
\newcommand{\vc}{v_c}
\newcommand{\VKOSTYA}{V_K}
\newcommand{\delEF}{\delta_F}
\newcommand{\brc}{{\vec r}_c}
\newcommand{\br}{{\vec r}}
\newcommand{\omth}{\omega_{\rm th}}
\newcommand{\omegathres}{\omega_{\rm th}}
\newcommand{\mata}{ {\bf a}}

%
%
\title{
Fermi Edge Singularities in the Mesoscopic Regime: \\
II. Photo-absorption Spectra}

%
\author{Martina Hentschel}
\affiliation{Max-Planck Institute for Physics of Complex Systems,
N\" othnitzer Stra\ss e 38, Dresden, Germany}
\affiliation{
Department of Physics, Duke University, Box 90305, Durham, North Carolina 27708-0305, USA}
\affiliation{
Institut f\"ur Theoretische Physik, Universit\"at Regensburg, 93040
Regensburg, Germany}

\author{Denis Ullmo}
\affiliation{
Department of Physics, Duke University, Box 90305, Durham, North Carolina 27708-0305, USA}
\affiliation{CNRS; Universit\'e Paris-Sud; LPTMS UMR 8626, 91405 Orsay Cedex, France}

\author{Harold U. Baranger}
\affiliation{
Department of Physics, Duke University, Box 90305, Durham, North Carolina 27708-0305, USA}

\date{June 18, 2007}

%
%
\begin{abstract}
We study Fermi edge singularities in photo-absorption spectra of generic mesoscopic systems such as quantum dots or nanoparticles. We predict deviations from macroscopic-metallic behavior and propose experimental setups for the observation of these effects. The theory is based on the model of a localized, or rank one, perturbation caused by the (core) hole left behind after the photo-excitation of an electron into the conduction band. The photo-absorption spectra result from the competition between two many-body responses, Anderson's orthogonality catastrophe and the Mahan-Nozi\`{e}res-DeDominicis contribution. Both mechanisms depend on the system size through the number of particles and, more importantly, fluctuations produced by the coherence characteristic of mesoscopic samples. The latter lead to a modification of the dipole matrix element and trigger one of our key results: a rounded K-edge typically found in metals will turn into a (slightly) peaked edge on average in the mesoscopic regime. We consider in detail the effect of the ``bound state'' produced by the core hole.
\end{abstract}
\pacs{73.21.-b,78.70.Dm,05.45.Mt,78.67.-n}
\maketitle

\section{Introduction}
\label{sec_intro}


Fermi edge singularities (FES) are among the simplest many-body effects found in condensed matter physics, and have been studied extensively  for bulk systems \cite{mahan:book,nozieres,tanabe:RMP1990}.  The continuing progress in the fabrication and experimental investigation of mesoscopic systems makes probable that such singularities could be observable relatively soon in quantum dots or nanoparticles, for which interference effects, and thus mesoscopic fluctuations, have to be taken into account. The study of Fermi edge singularities in this mesoscopic regime is the topic of the present paper; it extends and deepens our previous results \cite{xrayprl,xrayaoc}.

For macroscopic systems, the physics we discuss is known as the ``x-ray edge problem'' and is well established and understood \cite{mahan:book,nozieres,tanabe:RMP1990}. In an x-ray absorption process, a core electron is excited into the conduction band. It leaves behind a positively charged hole that can be considered as a static impurity.  For electronic densities corresponding to an interaction parameter $r_s \!\sim\! 1$, as realized typically in both semiconductors and metals, the screening length is of order the Fermi wavelength. This impurity can therefore be considered as localized (point like) once screening by the conduction electrons is taken into account. A similar situation arises after the excitation of a valence electron in semiconductor photoluminescence studies in which recombination occurs at an impurity.

The conduction electrons respond to such an abrupt, non-adiabatic perturbation by slightly adjusting their single particle energies and wave-functions. 
Although the overlap of the single particle states before and after the perturbation is very close to one, the overlap $\Deltaov$ between the initial and final many-body ground states tends to zero in the thermodynamic limit as a power of the number $M$ of particles. This effect 
is known as the Anderson orthogonality catastrophe (AOC) \cite{anderson:PRL1967}. As a result of AOC, the photo-absorption cross section $A(\omega)$ will be power-law suppressed for energies $\omega$ near the threshold energy (Fermi edge) $\omegathres$.

In the x-ray edge problem, AOC competes with a second, counteracting many-body response, also related to screening of the impurity potential by the conduction electrons. It is often referred to as Mahan's exciton, Mahan's enhancement, or the Mahan-Nozi\`{e}res-DeDominicis (MND) contribution \cite{mahan:book,nozieres}. In contrast to AOC which acts universally, the MND response depends on fulfilling dipole selection rules. It therefore depends on the symmetry of the conduction and core states. We will distinguish two symmetries of the core electron wave-function throughout the paper: $s$-like symmetry (corresponding to the K-shell of the atom) and $p$-like symmetry ($L_{2,3}$-shell). The respective thresholds are known as the K- and L-edge.

\begin{figure}
\includegraphics[width=8.5cm]{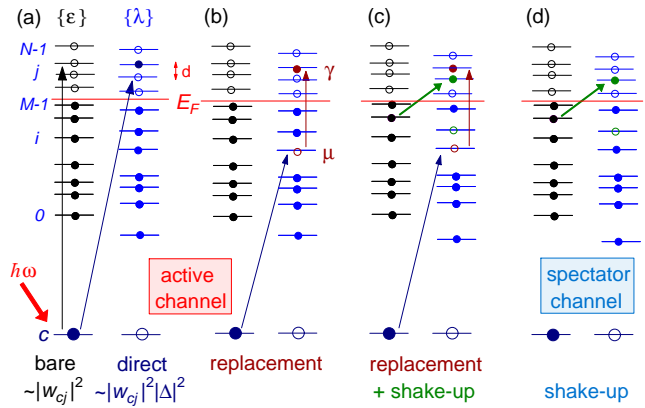}
\caption{(Color online) Schematic illustrating processes contributing to the photo-absorption cross section in a Fermi golden rule approach. Conduction electrons (mean level spacing $d$) in a generic chaotic system initially occupy levels $\eps_0 \ldots \eps_{M-1}$ (filled) and $\eps_M \ldots \eps_{N-1}$ (empty), that lower to $\lam_0 \ldots \lam_{N-1}$  when the core electron $c$ is excited. Optically active electrons contribute via the coherent superposition of (a) direct  and (b) replacement processes. In addition, one (or more) electron-hole pairs can be generated in shake-up processes, (c) for optically active channel and (d) for spectator channel, which are especially important away from threshold. Their presence reflects the suddenness of the perturbation. The vertical arrow in (a) is the bare process which represents the photo-absorption cross section in the naive picture without many-body effects. It depends only on the dipole matrix element $w_{c j}$ between the core electron $c$ and the single particle state $j$, $A^b(\omega) \propto |w_{c j}|^2$. The effect of AOC is accounted for in the direct process by the additional factor $|\Deltaov|^2 < 1$, $A^d(\omega) \propto |w_{c \gamma}|^2 \; |\Deltaov|^2$, 
  \label{fig_levelscheme}}
\end{figure}

The bulk x-ray edge problem was analyzed in detail using different techniques. Early approaches by Mahan\cite{mahan:book} and Nozi\`{e}res and co-workers\cite{nozieres} treated it based on field-theoretical methods and diagrammatic perturbation theory in analogy to the Kondo problem \cite{kondo_hewson}. Schotte and Schotte \cite{schotteschotte} used a Fermi golden rule approach and employed bosonization techniques for the rotationally invariant, effectively one-dimensional, scattering potential. 
The x-ray edge problem was also addressed by, among others, Friedel \cite{friedelscomment} and Hopfield \cite{hopfield}. In the 1980's Tanabe and Ohtaka demonstrated in detail the suitability of a Fermi golden rule approach \cite{tanabe:RMP1990,tanabe:seriesofpapers}. This method explicitly uses the fact that the impurity in the x-ray edge problem is static rather than dynamic, as in the Kondo problem. The spirit of the Fermi golden rule approach is illustrated schematically in Fig.~\ref{fig_levelscheme}; it is the approach used here.

The effect of a local perturbation such as the one induced by the core hole is characterized by the partial-wave phase shifts $\delta_l$ (at the Fermi energy) for each orbital channel.  The phase shifts have to obey the Friedel sum rule $Z\!=\!\sum_l 2 (2 l+1) \delta_l/\pi$ with the screening charge $Z\!=\!-1$ in our case.  The factor of two accounts for spin. 

The result of all the bulk techniques is that the photo-absorption cross section $A(\omega)$ near threshold $\omegathres$ has the form
\begin{equation}
A(\omega) \propto (\omega-\omegathres)^{-2 \fr{|\delta_{l_o}|}{\pi}
+ \sum_l 2 (2 l+1) \li[ \fr{\delta_l}{\pi} \re]^2} \:.
\label{eq:edgeshape}
\end{equation}
The first term in the exponent involves only the optically active channel (labeled $l_0$) and is the MND contribution, whereas the second term sums over all channels and corresponds to AOC. Note the different functional dependence on the phase shifts, linear and quadratic, respectively.

We will, throughout this paper, assume that (i) the local part of the conduction electron's wave-function -- at the lattice level -- is featureless (i.e.~of $s$-type) and (ii) the perturbation created by the core hole is spherically symmetric. In the bulk, it follows from the first assumption that the optically active channel, for which core and conduction electrons are linked by the dipole operator, is $l_0 \!=\! 1$ for K-shell core electrons and $l_0 \!=\! 0$ for the L-shell. As a consequence, for the K-shell, the perturbation acts on a channel which is \textit{not} optically active.  Therefore, the absorption spectrum is only affected by AOC in the $l\!=\!0$ channel, yielding a suppression, or rounding, of the edge of the photo-absorption spectra (``rounded edge''). On the other hand, for L-shell core electrons, the optically active channel \textit{is} the one affected by the perturbation, and the corresponding materials typically show an enhancement of the photo-absorption at the threshold frequency (``peaked edge'').  For a detailed analysis of the x-ray spectra of bulk Li, Na, Mg, and Al, we refer the reader to Ref.\,\onlinecite{citrin:PRB1979}, where in addition effects due to phonon excitations, the finite lifetime of the hole, and the deviation from spherical symmetry of both the local part of the conduction electron state and the perturbing potential are considered, all of which we will neglect here in order to focus on the essential mesoscopic physics.

In semiconductors, typically only $s$-like conduction electrons exist which, consequently, have to provide all the screening of the core hole. From the Friedel sum rule we then find $\delta_0 \!=\! -\pi/2$. This implies that we are in the strong perturbation regime which is typically not realized in bulk metals or related nanoscale structures like metallic nanoparticles. We shall see below that this has significant physical consequences related to the formation of a bound state.

In contrast, in metals, electrons of all channels $\delta_l$ typically contribute to Friedel screening. Thus the phase shift in the optically active channel is not $\pi/2$. However, the x-ray edge physics can be successfully captured based on $\delta_0$ alone, i.e., by assuming a spherically symmetric core hole potential \cite{tanabe:RMP1990,nozieres,citrin:PRB1979,ohtaka:private}. This remains true in the mesoscopic regime. For this situation, then, one considers $\delta_0 \!<\! \pi/2$ even though there is only one phase shift in the model. Thus, to address both the semiconductor and metal situations, we include results for the full range $\delta_0 \!\le\! \pi/2$.

The above description applies to clean bulk systems. The question we would like to address in this paper is whether the results found for the edge behavior in metals also hold, or have to be modified, when considering smaller, fully coherent, mesoscopic or even nanoscopic samples. Here, we study the universal class of {\it chaotic} ballistic systems, to which a random matrix model of the energy levels and wave-functions applies.

Reducing the size of the system will affect the Fermi edge singularity in various ways. First of all, the number of particles (electrons) $M$ in the system will be finite -- we are not in the thermodynamic limit any longer -- and for instance the power law dependence of $\Deltaov(M)$ causes, in fact, a huge difference in the efficiency of AOC in systems with $M \!\sim\! 10^{23}$ from those with, say, $M \!\sim\! 100$ electrons. Secondly, as the system becomes fully coherent, a plane wave (Bloch wave) description of the conduction electrons does not apply any longer. Rather, we need the actual wave-function of the specific mesoscopic systems to be described.  One consequence is that, since the confining potential will in general destroy spherical symmetry, angular momentum is lost as a quantum number, and the dipole selections rules need to be modified.  Furthermore, interference effects, and therefore mesoscopic fluctuations, need to be taken into account. All of these affect both the Anderson orthogonality catastrophe and the Mahan-Nozi\`{e}res-DeDominicis contribution.

Aspects of AOC in {\it disordered} mesoscopic systems have been addressed in Refs.\,\onlinecite{vallejos:PRB2002} and~~\onlinecite{gefen:PRBR2002}. AOC and x-ray photoemission spectra (where the excited core electron leaves the metal and the edge behavior is determined by the AOC response alone) were studied in Ref.\,\onlinecite{kroha:PRB1992} for impure simple metals. AOC in {\it ballistic chaotic} systems was the subject of the first paper\cite{xrayaoc} in this series.  Our main findings -- in line with the results in Refs.\,\onlinecite{vallejos:PRB2002} and~~\onlinecite{gefen:PRBR2002} -- are (i) incompleteness of AOC due to the finite number of particles, and (ii) a broad distribution of AOC-overlaps as a result of mesoscopic fluctuations which (iii) are dominated by the levels around the Fermi energy.

For (mesoscopic) x-ray absorption or photoluminescence spectra both the AOC and MND effects are of importance.  The MND response will, of course, also depend on the system size.  However, the relative strength of the two competing many-body responses might change as the system shrinks.  Optical FES in one-dimensional quantum wire systems have been studied both in experiment \cite{calleja} and theory \cite{oreg:PRB1996}. In contrast, FES in the photo-absorption spectra of two- or three-dimensional mesoscopic system have, to the best of our knowledge, not been addressed in literature.

The present paper aims at filling this gap. It is organized as follows. In Sec.~\ref{sec_model} we introduce our model for the (rank one) perturbation of chaotic conduction electrons described by random matrix theory, the dipole matrix element, and how spin is taken into account.  In Sec.~\ref{sec_boundstate} we consider in more detail the formation and role of the bound state appearing for strong perturbations.  In Sec.~\ref{sec_method} we explain our (Fermi golden rule) approach to the photo-absorption cross section.  The results are presented in Sec.~\ref{sec_results_avg} for the average photo-absorption cross sections at the K- and L-edges, and in Sec.~\ref{sec_results_fluct} for the mesoscopic fluctuations.  We devote Sec.~\ref{sec_exp} to the discussion of feasible experimental setups that would allow probing of our results, and close with a summary in Sec.~\ref{sec_summary}.

\section{Model}
\label{sec_model}

\subsection{Initial and Final Hamiltonian}

In our model of a quantum dot or nanoparticle, the electrons are confined in a coherent, irregularly shaped system. We describe the unperturbed system (the conduction electrons before a core electron is excited) by the Hamiltonian 
\be
\HamO = \sum_{k , \sigma} \eps_k c^\dagger_{k,\sigma} c_{k,\sigma}
\label{eq:hamH0}
\ee
with discrete eigen-energies $\eps_k$ ($k=0,\ldots,N-1$). The operator $c^\dagger_{k,\sigma}$ creates a particle with spin $\sigma = \uparrow, \downarrow $ in the orbital $\varphi_k(\br)$. The energy levels of the unperturbed system follow the statistics of random matrix theory\cite{mehta,bohigas} (RMT) and are characterized by a mean level spacing $d$, cf.~Fig.~\ref{fig_levelscheme}. We will distinguish situations where time-reversal symmetry is present (circular orthogonal ensemble, COE) from those where it is broken by, e.g., the presence of a magnetic field (circular unitary ensemble, CUE) \cite{mehta,bohigas}. As the number of electrons does not change in the processes that we consider, we drop the charging energy term normally present in isolated mesoscopic systems. Furthermore, we neglect any change in the residual electron-electron interactions \cite{UllmoBaranger01}.

This initial situation is perturbed  by a  rank one  or  contact potential $\HamV$ acting  at the location  $\brc$ of the core electron \cite{tanabe:RMP1990,mahan:book,nozieres}. Models for perturbations which are more general than rank one can also be  considered \cite{smolyarenko:PRL2002}.  It is necessary to consider them, for instance, for high density electron gas ($r_s \!\ll\! 1$) for which  the screening length  is significantly larger than the Fermi wavelength. For the density range corresponding to $r_s \!\sim\! 1$ which we consider here (as realized typically in both semiconductors and metals), the rank one approximation is, however, appropriate. The strength of the interaction between the core  hole and the conduction electrons is quantified by the  parameter $\vc <0$, which in turn  is related to the phase shift in the band center, $\delta_0$, by \cite{matveev:PRL1998,xrayaoc}
\begin{equation}
  \delta_0 = \arctan \frac{ \pi \vc}{d} \:.
  \label{eq:delta0}
\end{equation}
In addition to $\vc$, the effectiveness of the perturbation depends also on the wave-functions' amplitude $\varphi_k(\brc)$ at the position of the perturbation, which, in a mesoscopic system, will vary from state to state. In terms of $u_k \equiv \sqrt{\Omega} \, \varphi_k(\brc) $ so that $ \langle| u_k|^2 \rangle = 1$ (with $\Omega$ denoting the volume in which the electrons are confined),  the perturbation can be expressed as
\begin{equation}
\HamV = \vc \sum_{kk'} u_k^* u_{k'} c^\dagger_k c_{k'} \; .
  \label{eq:Vc}
\end{equation}

For reasons of comparison, we also define the {\it bulk-like} situation:
equidistant unperturbed energy levels a distance $d$ apart and
constant $u_k \equiv 1$ throughout the sample.  More details concerning this
model are given in Sec.~II of Ref.\,\onlinecite{xrayaoc} (the first
paper of the series).

Introducing $\tilde c^\dagger_{\kappa,\sigma}$ as creator of a particle in the perturbed orbital $\psi_{\kappa}(\br)$, we obtain the diagonal form of the perturbed Hamiltonian
\be
\Ham = \HamO + \HamV = \sum_{\kappa , \sigma} \lambda_{\kappa}
\tilde c^\dagger_{\kappa,\sigma} \tilde c_{\kappa,\sigma}\:.
\label{eq:ham}
\ee
(Note that we will use Greek letters to refer to the perturbed system.) In analogy to $u_k$, we refer to the perturbed amplitudes by $\tilde u_{\kappa} \equiv \sqrt{\Omega} \, \psi_{\kappa} (\brc) $. Finally, we define the transformation matrix ${\bf{a}} = (a_{ \kappa k})$,
\begin{equation}
\psi_{\kappa} = \sum_{k=0}^{N-1} a_{\kappa k} \varphi_k \:,
\label{eq:intromata}
\end{equation}
for later use. For relations between $u_k,$ $ \tilde u_{\kappa}$, $a_{k \kappa}$, and the $\eps_k,  \lam_{\kappa} $ see Eqs.~(15)-(19) in Ref.\,\onlinecite{xrayaoc}.

We denote the $M$-particle, Slater-determinant ground states
of the unperturbed and the perturbed system by $| \Phi_0 \ra$ and $|\Psi_0 \ra$, respectively. Their overlap is the Anderson overlap $\Deltaov$ which we considered in detail in Ref.\,\onlinecite{xrayaoc}. In the case of a rank one perturbation, it can be expressed as a function of the unperturbed and perturbed energy levels alone\cite{tanabe:RMP1990},
\begin{equation}
|\Deltaov|^2=
|\la \Psi_0 | \Phi_0 \ra |^2 = \prod_{i=0}^{M-1}  \prod_{j=M}^{N-1}
\fr{ (\lam_j - \eps_i)  (\eps_j - \lam_i) }{ (\lam_j - \lam_i)  (\eps_j - \eps_i)} \:.
\label{eq:overlap}
\end{equation}
Note that whenever possible, we use the index $j$ for levels above $E_F$ ($\gamma$ for perturbed levels), and $i$ ($\mu$) for levels below $E_F$. Furthermore the index $k$ ($\kappa$) is reserved for reference to all unperturbed (perturbed) levels.

\subsection{Dipole Matrix Element}
\label{sec_dipole}

The dipole operator is $\hat{D} = (e E_0/c\,m) (\vec{e} \cdot \vec{p} + h.c.)$ where $E_0$ is the magnitude of the electric field and $\vec{e}$ is its polarization. 
The dipole matrix element
\be w_{c \mu} \df \la \varphi_c |
\hat{D} | \psi_{\mu} \ra
\label{eq:wcmu}
\ee
is therefore proportional to the overlap of the perturbed wave-function $\psi_{\mu} $ with the derivative along $\vec{e}$ of the core electron wave-function $|\varphi_c \ra$.  For K-shell core electrons (spherically symmetric), $\varphi_c$ is, on the scale of the Fermi wavelength of the conduction electrons, essentially a $\delta$-function. In that case, $w_{c \mu} $ is  proportional to the derivative of $\psi_{\mu}$ along $\vec{e}$.  For L-shell core electrons on the other hand, the derivative of $|\varphi_c \ra$ is approximately a $\delta$-function, and therefore $w_{c \mu}$ is in this case proportional to $\psi_{\mu}$ itself rather than to its derivative. As a consequence, one has
\be
w_{c \mu} \propto \left\{ \begin{array}{l}
                \psi_{\mu}'(\brc) \equiv \tilde{u}'_{\mu}
                        \quad \textrm{K-edge} \\[1mm]
                \psi_{\mu} (\brc) \equiv \tilde{u}_{\mu}
                        \quad  \textrm{L-edge}\:.
              \end{array}
     \right.
\label{eq:dipolepsipsiprchaotic}
\ee

In the bulk, this implies precisely the selection rules mentioned in the introduction, namely that the optically active channel (with non-zero $w_{c \mu}$) is $l\!=\!1$ for K-shell core electrons and $l\!=\!0$ for L-shell core electrons.  Since only the $l\!=\!0$ channel is affected by the rank-one perturbation in Eq.~(\ref{eq:Vc}), there is no MND enhancement for the K-shell, and thus it has a rounded Fermi edge due to AOC.

In the mesoscopic case, however, angular symmetry is usually broken by the confining potential. Assuming chaotic classical dynamics, the magnitude of the unperturbed wave-functions, $|\phi_{\mu}|^2$, and the corresponding derivatives, $|\phi_{\mu}'|^2$, 
at a given position are statistically independent and both obey the Porter-Thomas distribution \cite{bohigas,prigodin}. The perturbed wave-functions and their derivatives can then be found from the transformation Eq.~(\ref{eq:intromata}). Thus, in a mesoscopic situation, the MND response does \textit{not} vanish even at the K-edge. We shall see below that this indeed leads to qualitative differences in comparison with the bulk behavior: We predict a slightly peaked, rather than rounded, K-edge in generic nanosystems. At the L-edge, there is a strong MND response, similar to that in the bulk metallic case, because the dipole matrix element is directly proportional to the amplitude of the perturbed wave-function at the position of the perturbation.

\section{Bound State}
\label{sec_boundstate}

\begin{figure} 
\includegraphics[width=8.5cm]{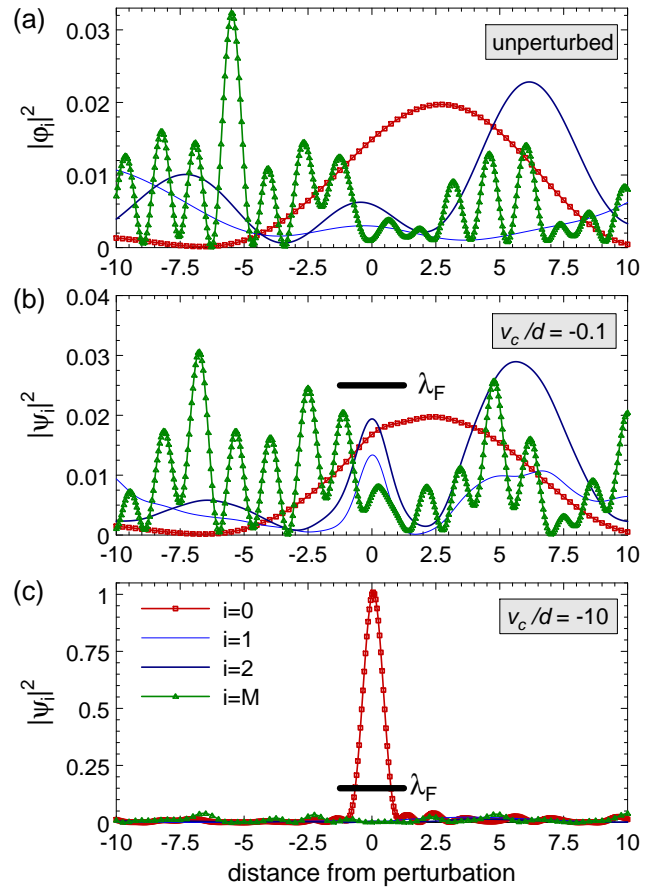}
\caption{(Color online) Chaotic single-particle wave-functions subject to a rank one perturbation.  (a) Unperturbed and (b), (c) perturbed wave-function probabilities for $i=0,1,2$ and $M$, i.e., for the lowest three eigenstates and the state at the Fermi energy $E_F$.  We assume a two-dimensional chaotic system ($N\!=\!100, M\!=\!50$) with unperturbed energies at their mean (bulk-like) values and model the unperturbed wave-functions as random superpositions of $100$ plane waves. The intensities along a line that contains the perturbation, located at $\brc \!=\! 0$, are shown.  The normalization volume of the wave-function is $\Omega \!=\! N $.
In (b), a weak perturbation causes only slight changes in the intensities.  In contrast, a strong perturbation in (c) causes the wave-function intensity $|\psi_0|^2$ corresponding to the bound state to pile up at the position of the perturbation. Screening of the core hole is done by the bound state on a length scale of order the Fermi wavelength, $\lam_F$, indicated by the black bar.
 \label{fig_rank1pileup}}
\end{figure}

When the perturbation strength exceeds a certain value, approximately $\vc/d \leq -1$, the lowest perturbed energy level $\lam_0$ will have an energy significantly below all the other levels (Fig.~\ref{fig_levelscheme}).  As discussed in  Ref.\,\onlinecite{xrayaoc} the average position  $\lam_0$ is given in this regime by
\be
    \lam_0 - \eps_0 = -\frac{N \Deltaml}{e^{d/|\vc|} - 1}\:,
\ee
and its fluctuations are negligible.

As discussed by Friedel \cite{friedel_boundstate} and others \cite{combescotnozieres,zagoskin,tanabe:RMP1990}, the existence of this low energy level is associated with the formation of a \textit{bound state}, which completely screens the core hole perturbation potential.  To illustrate this we employ the Berry-Voros conjecture \cite{berry_planewave,voros_planewave} to model the unperturbed wave-functions of the chaotic mesoscopic system as a random superposition of plane waves. Figure \ref{fig_rank1pileup}(a) shows wave-function intensities for both the low-lying eigenstates and those at the Fermi level. The final wave function intensities are then shown upon application of a weak [Fig.~\ref{fig_rank1pileup}(b)] or strong [Fig.~\ref{fig_rank1pileup}(c)] rank one perturbation. Whereas the small perturbation has little effect on the wave-function probabilities, the situation changes strikingly when a strong perturbation is applied: the lowest perturbed eigenstate $\psi_0$ concentrates at the position of the perturbation. The extension of the peak in $|\psi_0|^2$ is of order the Fermi wavelength $\lam_F$, which for systems with electronic densities corresponding to $r_s \geq 1$, to which our rank one perturbation model applies, is of order the screening length $\lam_{\rm screen}$.

The existence of a bound state has several important consequences. The first is the existence of a secondary band of absorption corresponding to final states for which the bound state is unoccupied (formation of a shake-up pair involving the bound state). The secondary band is well separated in energy from the main band (where the bound state is filled in the final state), and we shall not consider it here in much detail.

Another consequence of direct relevance to our study is the effect of the bound state on the magnitude of the dipole matrix elements because of the large amplitude $\psi_0(\brc)$. For strong perturbation, the probability density $|\psi_0(\brc)|^2$ at $\brc$ is significantly larger than the sum for all the other states (we return to this point below).  In contrast, this is not the case for the derivative of $\psi_0$ at $\brc$, since the largeness of the amplitude is compensated by the fact that $\psi_0$ possesses a maximum in the vicinity of $\brc$. Therefore, even for strong interaction, the bound state is not going to play a dominant role for the K-edge absorption spectra. On the other hand, for the L-edge, where the dipole matrix element is proportional to $\psi_\kappa$,  replacement processes through the bound state will, for strong perturbation, dominate the absorption.

In light of this discussion of the bound state, it is instructive to see how the Friedel sum rule is treated in our model Hamiltonian approach, Eqs.~(\ref{eq:hamH0})-(\ref{eq:ham}), in the semiconductor case in which there is a single type of conduction electron. Indeed, in this model, the Coulomb interaction between electrons is neglected, but the fact that the charge of the core hole is $+e$ is taken into account by requiring that the phase shift at the Fermi energy caused by $\HamV$ is $\delta_0 \!=\! - \pi/2$.  In this way each spin channel provides half a charge to screen the core hole.

Looking at the energy dependence of the phase shift for our model, one realizes, however, that the manner of this screening is somewhat unintuitive. The most natural initial supposition is that the phase shift decreases smoothly from $0$ at the bottom of the band (ie. little effect of the perturbation on states at the band edge) to $-\pi/2$ at the Fermi energy. However, what happens in practice is that the phase shift starts at $-\pi$ at the bottom of the band and \textit{increases} gradually to $-\pi/2$ at the Fermi energy. In other words, the bound state provides a charge $-2e$ because both spin species have to be taken into account, which means that the core hole is actually \textit{over-screened}. All the other (perturbed) wave-functions, instead of participating in the screening, are actually pushed away from the location of the perturbation, providing an effective charge $+e$ near the core hole, leaving, of course, the required net screening charge of $-e$. 


A natural question is whether the scenario described above actually happens in physical (experimental) systems, where Coulomb interaction between the electrons and chemistry at the lattice level both occur. It is clear, for instance, that the two electrons occupying the bound state interact strongly with each other. If this interaction energy is larger than the difference between the bound state and Fermi energies, it will prevent double occupation of the bound state and give rise to a local moment and so Kondo physics. For the experimental realizations we have in mind, however, the bound state wave-function will be spread out on the scale of the Fermi wavelength (it is not a deep level of an impurity), and simple estimates show that a local moment is not expected.

Assuming Kondo physics is not involved, we now have to consider how ``real'' the bound state actually is in practice and if its properties are the same in experimental systems as in our model. This question is, of course, not specific to the mesoscopic problem.  As early as 1952, Friedel discussed in detail the physical reality of the bound state in rank one models \cite{friedel_boundstate} (see also the discussion by Combescot and Nozi\`{e}res in Ref.\,\onlinecite{combescotnozieres}). For example, the screening of the core hole at the lithium K-edge is done by the $2s$ conduction electrons, while the sodium L$_{2,3}$-edge is screened by its $3s$ conduction electrons \cite{friedel_boundstate}. It is precisely those $s$-orbitals which, according to the picture developed here following Ref.\,\onlinecite{friedel_boundstate}, take 
the role of a bound state.

Thus the boundstate is a physical reality. But one should bear in mind that its extension in space, which controls the size of the dipole matrix element $\omega_{c0}$, can be heavily influenced by the local chemistry or other factors not described by our model. In the two limiting cases -- (1)~if the absorption is entirely dominated by the bound state (L-shell with strong perturbation), or (2)~if the bound state is playing a negligible role (weak perturbation or K-shell) -- the fact that $\omega_{c0}$ may not have the proper physical value is of little relevance: In the former case, only an overall prefactor is involved in which we are in any case not interested, and in the latter case, an incorrect magnitude will obviously not affect at all the description. Note furthermore that in the strong perturbation limit, since the phase shift at the Fermi energy is essentially independent of $\vc$ for $\vc/d \ll -1$, $\vc$ can be chosen so that the energy of the bound state is the physical one.

The situation is more complicated for intermediate situations, where the processes involving the bound state have a similar contribution to the absorption spectra as those not involving the bound state (L-shell with $\vc/d \simeq -1$). Since our model has only one parameter, it cannot reproduce arbitrary values of both the phase shift at the Fermi energy and the ratio between $\omega_{c0}$ and the mean value of the other dipole matrix elements. In such a situation, it is then necessary to look into the details of the bound state for the physical system before employing our model. We shall come back to this point in Section~\ref{sec_exp} when discussing particular mesoscopic realizations. However, in the following, we mainly discuss the limiting cases for which the physical relevance is unambiguous.

\section{Method}
\label{sec_method}

We will see in this section that the absorption amplitude $A(\omega)$ for the L-edge is determined entirely by the unperturbed and perturbed eigenvalues $\{ \eps \}$ and $\{ \lam \}$. For the K-edge, knowledge of the derivative of the unperturbed wave-functions at $r_c$ is also required. For a chaotic system, these latter are statistically independent of the spectra and obey a Porter-Thomas distribution \cite{prigodin}, and thus do not pose any particular difficulty. To study numerically the (statistical) properties of $A(\omega)$, we therefore generate one particle spectra with the known joint distribution \cite{matveev:PRL1998} by using a Metropolis algorithm and deduce from them all the statistical quantities needed to characterize the absorption spectra. (Some details about the algorithm we have used are given in footnote\,\,\onlinecite{metropolis}.)  This Section derives the basic expressions needed for this purpose.

{\it Photo-absorption cross section.}---  Following the Fermi
golden rule based treatment by Tanabe and Ohtaka\cite{tanabe:RMP1990}, we write the photo-absorption cross section $A(\omega)$ in terms of the matrix element of the dipole operator $\hat D$ between the unperturbed many-body ground state with the core level $c$ filled, $| \Phi_0^c \ra$ and energy $E_0^c$, and the perturbed (final) many-body state with an additional conduction electron $|\Psi_f \ra $ at energy  $E_f = E_0^c + \hbar \omega$,
\begin{equation}
  A(\omega) = \frac{2 \pi}{\hbar} \sum_f | \la \Psi_f | \hat D |
  \Phi_0^c \ra |^2 \delta(E_f - E_0^c - \hbar \omega) \:.
\label{eq:fgr}
\end{equation}

For clarity, we consider spinless electrons for now, and return at the end of this Section to the modification introduced by spins. The unperturbed ground-state, therefore, comprises $M$ electrons on levels $0$ to $M-1$ with the core level filled,
\be
|\Phi_0^c\ra = \prod_{i=0}^{M-1}
c^\dagger_{i} c^\dagger_c|0\ra \:.
\ee
The core electron is created (annihilated) by the operator $c^\dagger_c (c_c)$.
In the perturbed final states $|\Psi_f\ra$ there are $M\!+\!1$
conduction electrons and the core level is empty,
\be
|\Psi_f\ra =
\prod_{\kappa\,{\rm filled}}
\tilde{c}^\dagger_{\kappa} |0\ra \:.
\ee
The dipole operator in second quantized form is
\be
\hat D = \sum_{\kappa=0}^{N-1} w_{c \kappa} \li ( \tilde{c}^\dagger_{\kappa} c_c\; +\; {\rm h.c.} \re) \:;
\label{eq:dipoleoperator}
\ee
the dipole matrix elements $w_{c \kappa}$ were discussed in
Sec.~\ref{sec_dipole}.

{\it Photo-absorption processes at threshold.} -- Let us begin our discussion with the absorption at the threshold energy $\omth$. Then, the only possible final state is the perturbed ground state with the core electron excited to level $M$ just above the Fermi energy; no shake-up processes are possible. The direct process, Fig.~\ref{fig_levelscheme}(a), is defined by keeping only $\kappa\!=\!M$ in the dipole operator (\ref{eq:dipoleoperator}) -- the term which acts between the core electron and the lowest unfilled level. The contribution of the direct process is, then,
\bea
\lefteqn{ A^d({\omth}) }&&\\
& = &  \frac{2 \pi}{\hbar} \li | \la0| \tilde{c}_0  \tilde{c}_1 \ldots  \tilde{c}_M
                          | w_{c M}  \tilde{c}^\dagger_M c_c
                          | c^\dagger_{M-1}  \ldots c^\dagger_1 c^\dagger_0 c^\dagger_c | 0 \ra \re |^2
                   \nonumber \\
            & = &  \frac{2 \pi}{\hbar} |w_{c M}|^2 | \la \Psi_0 | \Phi_0 \ra |^2 \; \propto\; |w_{c M}|^2 |\Deltaov|^2
                   \nonumber \; .
\eea
Introducing for reference the amplitude for the bare process in which
many-body effects are ignored,  $ A^0({\omth}) \propto |w_{c M}|^2$,
the contribution of the direct process can be expressed as $A^d({\omth})
= A^0(\omth)  |\Deltaov|^2 $. This makes evident the  role of AOC and
the fact that $A^d$  vanishes in the thermodynamic limit.

Even at threshold the direct term is not the only contribution to the absorption amplitude: since $\tilde c_\kappa \neq c_\kappa$, the terms $\kappa < M$ in Eq.~(\ref{eq:dipoleoperator}) are non-zero. These are known as replacement processes, Fig.~\ref{fig_levelscheme}(b). In terms of the generalized overlap,
\be
\Deltaov_{\bar{\mu} \gamma} \equiv
\la \Psi_0 | \tilde{c}^\dagger_\mu \tilde{c}_\gamma  |\Phi_0 \ra \;,
\ee
of the unperturbed ground-state with the
perturbed state in which the particle in the orbital $\mu \leq M\!-\!1$ has been promoted to the orbital $\gamma \geq M$, the total photo-absorption cross section at threshold reads
\bea
\lefteqn{ A^{d,r}(\omth) }  \nonumber \\
& = &  \frac{2 \pi}{\hbar} \bigg| \Big\la0 \Big| \prod_{\mu=0}^M \tilde{c}_\mu
             \sum_{\mu=0}^{M} w_{c \mu}  \tilde{c}^\dagger_\mu c_c
             \prod_{i=M-1}^0 c^\dagger_{i}  \; c^\dagger_c \Big| 0 \Big\ra \bigg|^2
                   \nonumber \\
& = &  \frac{2 \pi}{\hbar} \bigg | w_{c M} \Deltaov
     \bigg ( 1 - \sum_{\mu=0}^{M-1}
     \frac{ w_{c \mu } \Deltaov_{\bar{\mu} M}} { w_{c M} \Deltaov }
                 \bigg ) \bigg |^2 \:.
\label{eq:dirreplthres}
\eea

{\it Direct and replacement contribution away from threshold.}--- For higher photon energies, such that the final state is obtained by adding a particle in the orbital $\gamma > M$ to the perturbed ground state $\Psi_0$, the last equation is readily generalized by the substitution $M \rightarrow \gamma$,
\be
 A^{d,r}(\omega) = \frac{2 \pi}{\hbar} \bigg | w_{c \gamma} \Deltaov
                 \bigg ( 1 - \sum_{\mu=0}^{M-1}
\frac{ w_{c \mu } \Deltaov_{\bar{\mu} \gamma}} { w_{c \gamma} \Deltaov }
                 \bigg ) \bigg |^2 \:.
\label{eq:dirrepl}
\ee
The replacement overlap enters the photo-absorption cross section via the ratio $w_{c \mu} \Deltaov_{\bar{\mu} \gamma} / ( w_{c \gamma} \Deltaov )$ in the second term. This ratio can be expressed as a product of eigenvalue differences using Eqs.\,(15)-(19) of Ref.\,\onlinecite{xrayaoc}. To this end, we need the dipole matrix elements $w_{c\mu}$ discussed in Sec.\,\ref{sec_dipole}. Recalling that $w_{c \mu} \!=\! \psi'_{\mu} (\brc) \!\equiv\! \tilde{u}'_{\mu}$ at the K-edge, and $w_{c \mu} \!=\! \psi_{\mu} (\brc) \!\equiv\! \tilde{u}_{\mu}$ at the L-edge, we express the ratio in Eq.~(\ref{eq:dirrepl}) as
\be
\frac{ w_{c \mu } \Deltaov_{\bar{\mu}  \gamma}} { w_{c \gamma} \Deltaov }
\equiv
{\cal{F}}_{\mu}
\frac{ \tilde{u}_{ \mu } \Deltaov_{\bar{\mu} \gamma}} { \tilde{u}_{\gamma} \Deltaov }
=
{\cal{F}}_{\mu}
\prod_{\scriptstyle{\kappa=0} \atop \scriptstyle{\kappa \neq \mu}}^{M-1}
\fr{\lam_{\kappa} - \lam_{\gamma} } {\lam_{\kappa} - \lam_{\mu}}
 \: \prod_{k=0}^{M-1} \fr{\eps_{k} - \lam_{\mu} } {\eps_{k} - \lam_{\gamma}}\:,
\label{eq:ratioforrepl}
\ee
where
\be
{\cal{F}}_{\mu} \equiv \left \{
                   \begin{array}{r}
                   \tilde{u}_{\gamma} \tilde{u}'_{\mu}   /
                    \tilde{u}_{\mu}   \tilde{u}'_{\gamma} 
                    \quad \quad \textrm{K-edge} \\[3mm]
                    1  \quad \quad \textrm{L-edge}
                   \end{array}
                  \right.
                  \label{eq:faccalf}
\ee
carries the symmetry-dependence of the dipole matrix elements.
The structure of Eq.~(\ref{eq:dirrepl}) suggests furthermore the
introduction of a function $p(\gamma)$,
\be
p(\gamma) \equiv w_{c \gamma}
\bigg( 1 - \sum_{\mu=0}^{M-1}
\frac{ w_{c \mu } \Deltaov_{\bar{\mu}  \gamma}} { w_{c \gamma} \Deltaov } \bigg) \:,
\ee
which can be computed using (\ref{eq:ratioforrepl}).

{\it Shake-up processes.}---
When the energy of the incident photon is at least two level spacings above the threshold energy, particle-hole pairs can be created in the final state, Fig.~\ref{fig_levelscheme}(c). In the one-pair shake-up contribution, two electrons are excited above $E_F$ in levels $\gamma_1$ and  $\gamma_2$ with the level $\mu_2$ empty in the final state. This situation can be handled in close analogy to the replacement process above by introducing a renumbered level sequence: let $\{\lam^1_{\kappa}\}$ be a renumbering of
$\{\lam_{\kappa}\}$ in which the level $\lam_{\mu_2}$ is skipped whereas $\lam_{\gamma_1}$ and $\lam_{\gamma_2}$ are appended as elements $\lam^1_{M-1}$ and $\lam^1_{M}$. Let $\{\tilde{u}^1\}$ and $\{ \tilde{u}^{\prime 1}\}$ be similarly renumbered sequences. Then, the
the one-pair shake-up photo-absorption cross section is \cite{tanabe:RMP1990}
\be
A^{\rm sh}(\omega)  =   \frac{2 \pi}{\hbar}
       \sum_{ \li. \{\mu_2,\gamma_1,\gamma_2 \} \re|_{\omega}} \bigg |
       \Deltaov_{\bar{\mu}_2 \gamma_1} \; p^1(\gamma_2) \bigg|^2 \:,
\label{eq:shuponepair}
\ee
where
\be
p^1(\gamma_2)  \equiv  w_{c \gamma_2}
\sum_{\mu=0}^{M-1}
\bigg[ {\cal{F}}^1_{\mu}
\prod_{\kappa=0 (\neq \mu)}^{M-1}
\fr{\lam^1_{\kappa} - \lam^1_{M} } {\lam^1_{\kappa} - \lam^1_{\mu}}
 \: \prod_{k=0}^{M-1} \fr{\eps_{k} - \lam^1_{\mu} } {\eps_{k} - \lam^1_{M}}
 \bigg]\:.
\ee
Here, the factor ${\cal{F}}^1_{\mu}$ generalizes Eq.~(\ref{eq:faccalf}) and takes the values
\be
{\cal{F}}^1_{\mu} = \left \{
                   \begin{array}{r}
            \tilde{u}^1_M  \tilde{u}^{\prime 1}_{\mu}  /
            \tilde{u}^1_{\mu} \tilde{u}^{\prime 1}_M 
                    \quad \quad \textrm{K-edge} \\[3mm]
                    1  \quad \quad \textrm{L-edge} \:.
                   \end{array}
                  \right.
\ee
Note that $A^{\rm sh}(\omega)$ does not change when $\gamma_1$ and $\gamma_2$ are interchanged in the above equations.

Generalization to cases with two and more shake-up pairs is straightforward: In Eq.~(\ref{eq:shuponepair}), the overlap of the initial state with a state with two or more shake-up pairs is needed. The corresponding functions $p^2$, $p^3,\ldots$ are obtained based on renumbering the energy levels such that an index shift occurs for each empty level below $E_F$, and the filled levels above are appended.

As the energy above the threshold increases, the number of energetically allowed final states with an arbitrary number of particle-hole excitations increases exponentially, and their exhaustive enumeration quickly becomes a hopeless task.  However, as we shall demonstrate in the next section, the number of final states {\em that actually contribute} to the absorption process remains finite, and actually not very large. This is what makes our approach practical in the end.

{\it Spin, the spectator channel.}--- We end this Section by discussing how the above picture is modified when the spin of the electrons is taken into account.  Let us choose the axis of spin quantization such that the core electron excited into the conduction band has spin up. Since the dipole operator $\hat{D}$ is spin independent, all the discussion above concerning direct, replacement and shake-up terms applies to the excited spin channel. The electrons with spin opposite to the excited spin -- referred to as the spectator channel -- are not connected by $\hat{D}$ to the core electrons; however, they are affected by the core hole potential, and their energies and wave-functions are modified.  As a consequence, the ground state is subject to the Anderson orthogonality catastrophe, and some excited states may have non-zero overlap with the unperturbed ground state, Fig.~\ref{fig_levelscheme}(d). Note that the energy of the incident photon can be shared between the two spin channels.

In practice, to account for the spin of the electrons the photo-absorption spectrum for the optically active channel has to be convoluted with curves 
for the spectator channel. This will lead to a slight smearing-out of features obtained for the photo-absorption cross section for the optically active channel, as we will see below.

\section{
The significant replacement and shake-up processes}

We have seen that it is straight forward to express the matrix elements appearing in the Fermi golden rule approach, Eq.~(\ref{eq:fgr}), in terms of the one particle energies $\{ \eps \}$ and $\{ \lambda \}$.  However, the total number of final states increases exponentially with the energy above threshold. It is therefore necessary to identify more precisely which final states do actually contribute to absorption, and show that the number of such states is not prohibitive.

\begin{figure}[b]
\hspace*{-3mm}
\includegraphics[width=8.5cm]{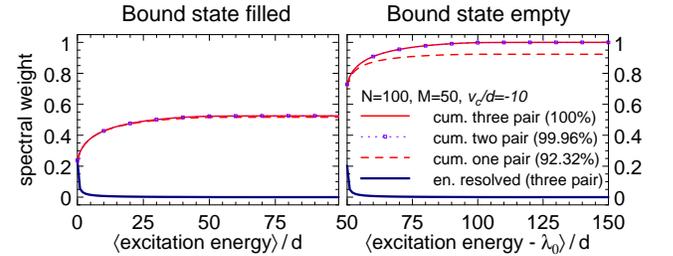}
\caption{(Color online) Spectral weight of the unperturbed many-body ground state $|\Phi_0\ra$ in terms of perturbed many-body states $|\Psi_{f}(\omega)\ra$, classified by their (average) energy from threshold
measured in units of mean level spacings $d$.
($\vc/d \!=\! -10, M\!=\!50, N\!=\!100$, COE statistics.)
The threshold for the secondary band (right panel), when the bound state is empty in the final state, is $M d$ greater than the threshold with bound state filled (left). The lower (thick blue) curve is the energy-resolved width of $|\Phi_0\ra$ in the perturbed basis, equivalent to the contribution of the spectator channel to the absorption cross section. The upper (thinner red) curves show the cumulative spectral weight taking into account terms with one, up to two, and up to three shake-up pairs (dashed, dotted with symbols, and full lines, respectively). Remarkably, less than 0.1\% of the weight is missed when including only up to two shake-up pairs. The slow saturation of the total weight (taking place on the scale of the band width) is a characteristic of AOC.
\label{fig_GSwidth}}
\end{figure}

Toward this end, Fig.\,\ref{fig_GSwidth} shows the spectral weight of the unperturbed (many-body) ground state in the perturbed basis, as a function of the perturbed state's energy. In addition, the cumulative spectral weight (summation up to the perturbed state energy) is shown. Including all possible terms and all energies, the cumulative spectral weight will, of course, be $|\Phi_0|^2 \!=\!1$. Fig.\,\ref{fig_GSwidth} tells us that in practice {\it not all} but rather {\it only a few} terms are needed to reach a spectral weight of $1$. In particular, it is not necessary to include shake-up processes of all orders and energies: Fig.\,\ref{fig_GSwidth} suggests that including terms with up to three shake-up pairs and energies up to one and a half the band width is  100\% sufficient and including terms with up to two shake-up pairs captures more than 99.9\% of the weight. Indeed, one-pair shake-up processes provide the dominant contribution to the spectral weight \cite{onepairshup_dominate} (about $92\%$ for $v_c \!=\! -10$). The calculation of the photo-absorption spectra below includes processes involving up to two shake-up pairs.

The thicker blue curves in Fig.~\ref{fig_GSwidth} show the energy distribution of the spectral weight (processes with up to three shake-up pairs are included). Evidently, a significant percentage of the total weight is borne by states of low energy (near the excitation threshold), or by states with energy equal to or slightly above the energy $\tilde E_{\rm bs} \!=\! \lambda_{M+1} \!-\! \lambda_0$ necessary to promote the bound sate electron into an empty orbital.

Because the dipole operator $\hat D$ is a one-particle operator, a non-negligible matrix element $\la \Psi_f | \hat D c^\dagger_c | \Phi_0 \ra$ requires $ | \Psi_f \ra \!=\! \tilde{c}^\dagger_\kappa | \Psi^0_f \ra$ where $\kappa$ can be arbitrary but $| \Psi^0_f \ra$ is restricted to be one of the states which overlaps significantly with $\Phi_0$. Therefore, the total number of final states that need to be considered grows only \textit{quadratically} with the energy above threshold, and there is no problem in enumerating them all.

\begin{figure}[b]
\includegraphics[width=8.5cm]{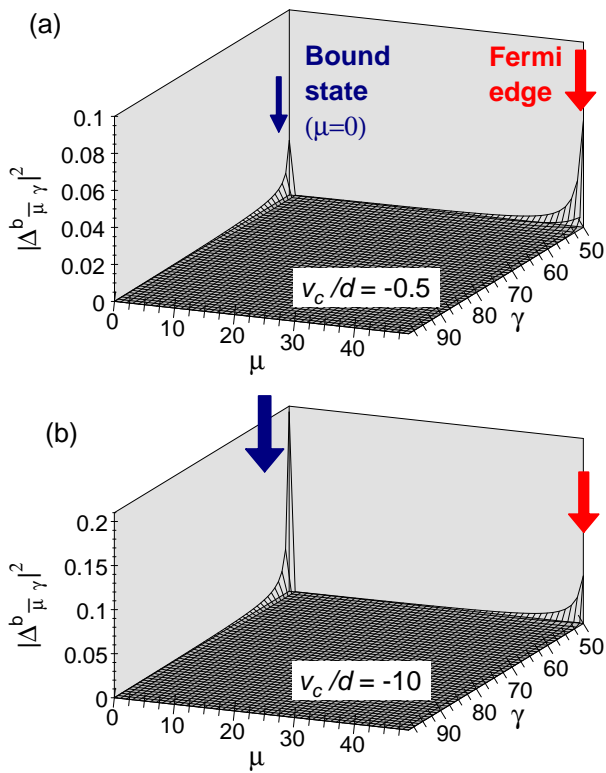}
\caption{(Color online) One-pair shake-up and replacement overlaps for the bulk-like case. (a) Intermediate perturbation strength, $v_c/d \!=\!-0.5$. (b) Strong perturbation, $v_c/d \!=\!-10$. For almost all $(\mu,\gamma)$ the replacement overlap $|\Deltaov_{\bar{\mu} \gamma}^b|^2$ is zero. Non-zero values arise for (1) replacement through the bound state with the excited electron close to $E_F$ ($\mu\!=\!0$ and $\gamma \!\gtrsim\! M$, black arrow), and (2) shake-up pairs formed in the vicinity of the Fermi edge ($\mu, \gamma$ both close to $M$, red arrow). For a strong perturbation, replacement through the bound state becomes the dominant process. \label{fig_reploverlap}}
\end{figure}

For processes involving only one shake-up pair, a more detailed representation of the decomposition of the unperturbed ground state is shown in Fig.~\ref{fig_reploverlap}, where $|\Deltaov_{\bar{\mu} \, \gamma}^b|^2$ is shown as a function of $\mu$ and $\gamma$ for the bulk-like case. These bulk-like overlaps, which also determine the replacement contribution, provide a good estimate for the mean value in the mesoscopic case and are useful to estimate the relative importance of the different processes. We see immediately that most of these overlaps are very small. There are two notable areas of exception: one where the particle-hole pair lies close to the Fermi energy ($\mu, \gamma \!\approx\! M$), and a second for terms involving the bound state. For a small perturbation, the Fermi energy peak dominates. However, as soon as a bound state develops upon increasing the perturbation, the overlaps $|\Deltaov_{\bar{0} \gamma}^b|^2$ start to grow and eventually overwhelm those near the Fermi energy.

\section{Photo-Absorption Spectra: Averaged Cross Section}
\label{sec_results_avg}


In the mesoscopic case, the photo-absorption threshold energy fluctuates from sample to sample. We assume here that $\omth$ can be determined experimentally, and that energies and spectra are then measured with respect to this energy.  Therefore, we will often give the photon excess energy relative to the threshold energy in mean level spacings, $\hbar \, (\omega \!-\! \omegathres)/d$.

Furthermore, in a mesoscopic system the final state energy $E_f$ can only take discrete values. A mesoscopic photo-absorption spectrum is therefore comprised of a series of $\delta$-peaks (broadened in experiment). We are interested in particular in the prefactor given by the Fermi golden rule matrix elements. First, we discuss their average in this Section, turning to fluctuations in the next. In all cases, we present only the photo-absorption from the primary band, for which the bound state is occupied in the final state. The absorption is normalized to the total absorption of the primary band.

\subsection{K-Edge}
\label{sec_result_avg_K}

\begin{figure}[t]
\includegraphics[width=8.5cm]{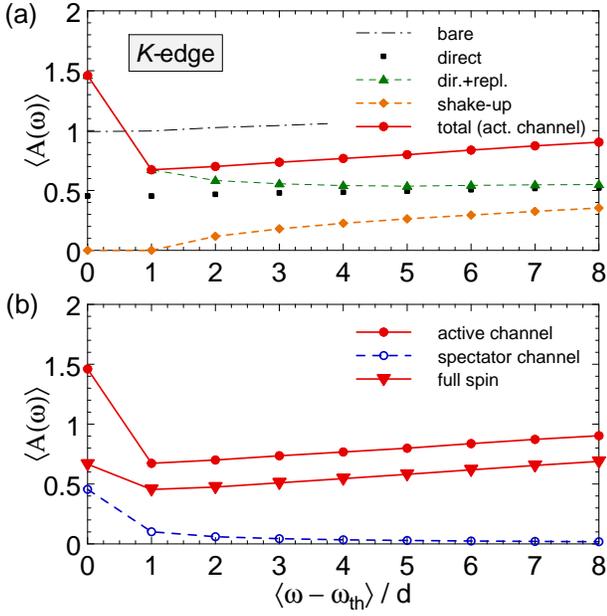}
\caption{(Color online) Average mesoscopic absorption spectra at the K-edge ($N\!=\!100, M\!=\!50, \vc/\Deltaml\!=\!-10$, CUE). (a) The total absorption cross section in the active channel (full line) is the sum of the direct/replacement (triangles) and shake-up (diamonds) contributions. For comparison, the bare contribution (dashed-dotted line) and the direct process alone (squares) are shown. 
(b) Active (full circles) and spectator (open circles) channel spectra separately, as well as the full spin photo-absorption cross section obtained after convolution (down-triangles). The edge is slightly peaked.
  \label{fig_Kedge_avgspec}}
\end{figure}

Fig.\,\ref{fig_Kedge_avgspec} shows the average photo-absorption cross section for a K-edge.  The time-reversal non-invariant case (CUE) is considered [see Ref.\,\onlinecite{xrayprl} for a similar illustration of the time-reversal invariant case (COE)]. Note that the replacement contribution decreases rapidly away from threshold, at which point shake-up processes become important. \textit{The large replacement amplitude at threshold causes the mesoscopic K-edge to be peaked.} 

Fig.\,\ref{fig_Kedge_avgspec}(b) shows the modification made by spin. The contribution of electrons with spin opposite to the excited core electron -- the spectator channel -- is shown. The full spin cross section is obtained after convolution of the spectra of the two spin species. The slight peak at the K-edge threshold is maintained in the full spin spectrum.

\begin{figure}[t]
\includegraphics[width=8.5cm]{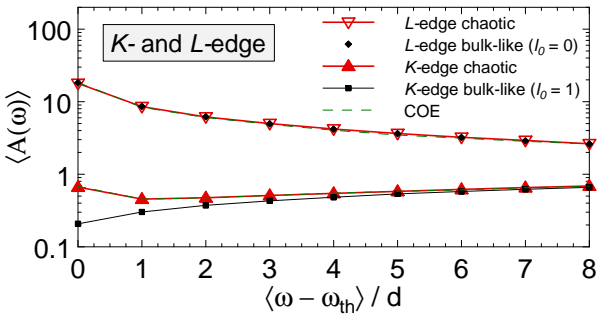}
\caption{(Color online) Average mesoscopic (triangles) and bulk-like (quadrangles) spectra as a function of energy from threshold at both a K- and L-edge ($N\!=\!100, M\!=\!50, \vc/d\!=\!-10$, CUE). Whereas the bulk-like and mesoscopic-chaotic results coincide for the L-edge, there is a clear difference in the K-edge spectra: the bulk-like edge is rounded whereas a generic mesoscopic system yields a slightly peaked edge on average. The dashed curves are the mesoscopic spectra in a COE situation; they are nearly indistinguishable from the CUE case.
  \label{fig_KLedge}}
\end{figure}

Comparison of the mesoscopic and bulk-like situations is shown in Fig.\,\ref{fig_KLedge}. At a K-edge (lower curves), the mesoscopic and bulk-like  photo-absorption spectra are qualitatively  different. The bulk-like K-edge shows the rounded behavior expected from AOC, though the threshold value is non-zero due to  incompleteness of AOC in a finite system. The average mesoscopic K-edge spectra, on the other hand, is slightly peaked at threshold. However, as the photon energy becomes only a few mean level spacings above threshold, the average mesoscopic K-edge photo-absorption quickly approaches the bulk-like result from above.

The reason for the different K-edge behavior in bulk-like and mesoscopic samples lies in the dipole matrix elements. In the bulk-like situation, dipole selection rules cause the matrix elements in the $l\!=\!0$ channel to vanish, and therefore there is zero MND response. In contrast, the dipole matrix elements in the mesoscopic situation are generally non-zero random numbers with a Porter-Thomas distribution, originating from the distribution of wave-function derivatives as discussed in Sec.~\ref{sec_dipole}. Therefore, there is a MND response near threshold in the mesoscopic situation that counteracts the AOC edge rounding, thus causing the peaked K-edge.

\subsection{L-Edge}
\label{sec_results_avg_L}

\begin{figure}[t]
\includegraphics[width=8.5cm]{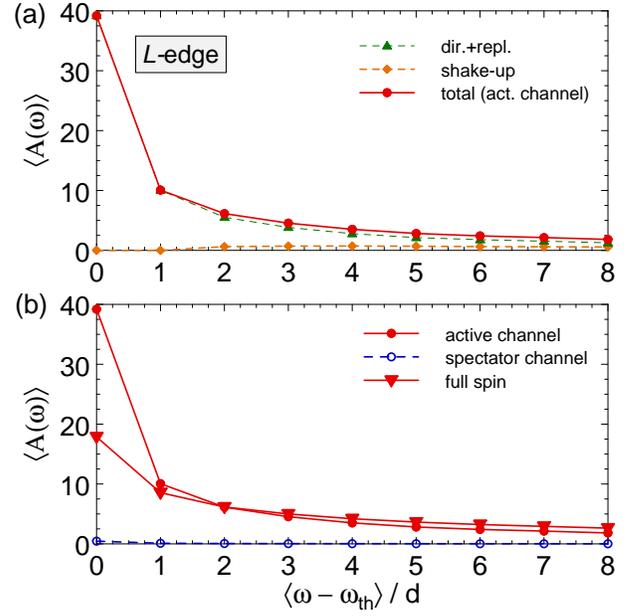}
\caption{(Color online) Average mesoscopic absorption spectra at the L-edge ($N\!=\!100, M\!=\!50, \vc/d\!=\!-10$, CUE). (a) The spectrum of the optically active channel is the sum of the direct/replacement (triangles) and shake-up (diamonds) contributions. (b) Active (full circles) and spectator (open circles) channel spectra separately, as well as the full spin photo-absorption cross section obtained after convolution (down-triangles). The contribution of the spectator channel is identical to that in Fig.~\ref{fig_Kedge_avgspec}(b). The peak at the L-edge is much more pronounced than that at the K-edge (Fig.\,\ref{fig_Kedge_avgspec}) and extends over several mean level spacings in photon energy.
  \label{fig_Ledge_avgspec}}
\end{figure}

\begin{figure*}
\includegraphics[width=8.5cm]{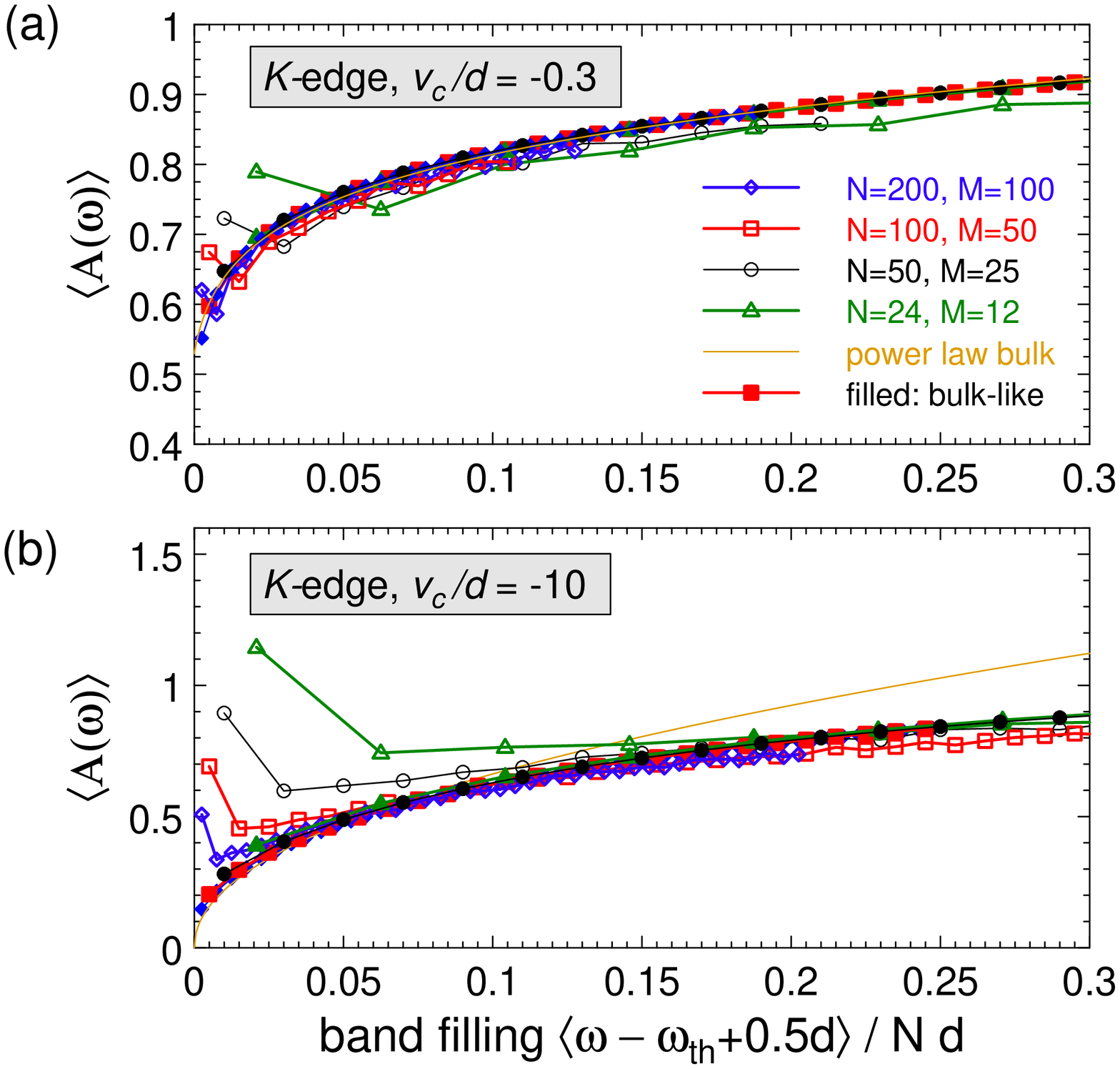}
\includegraphics[width=8.5cm]{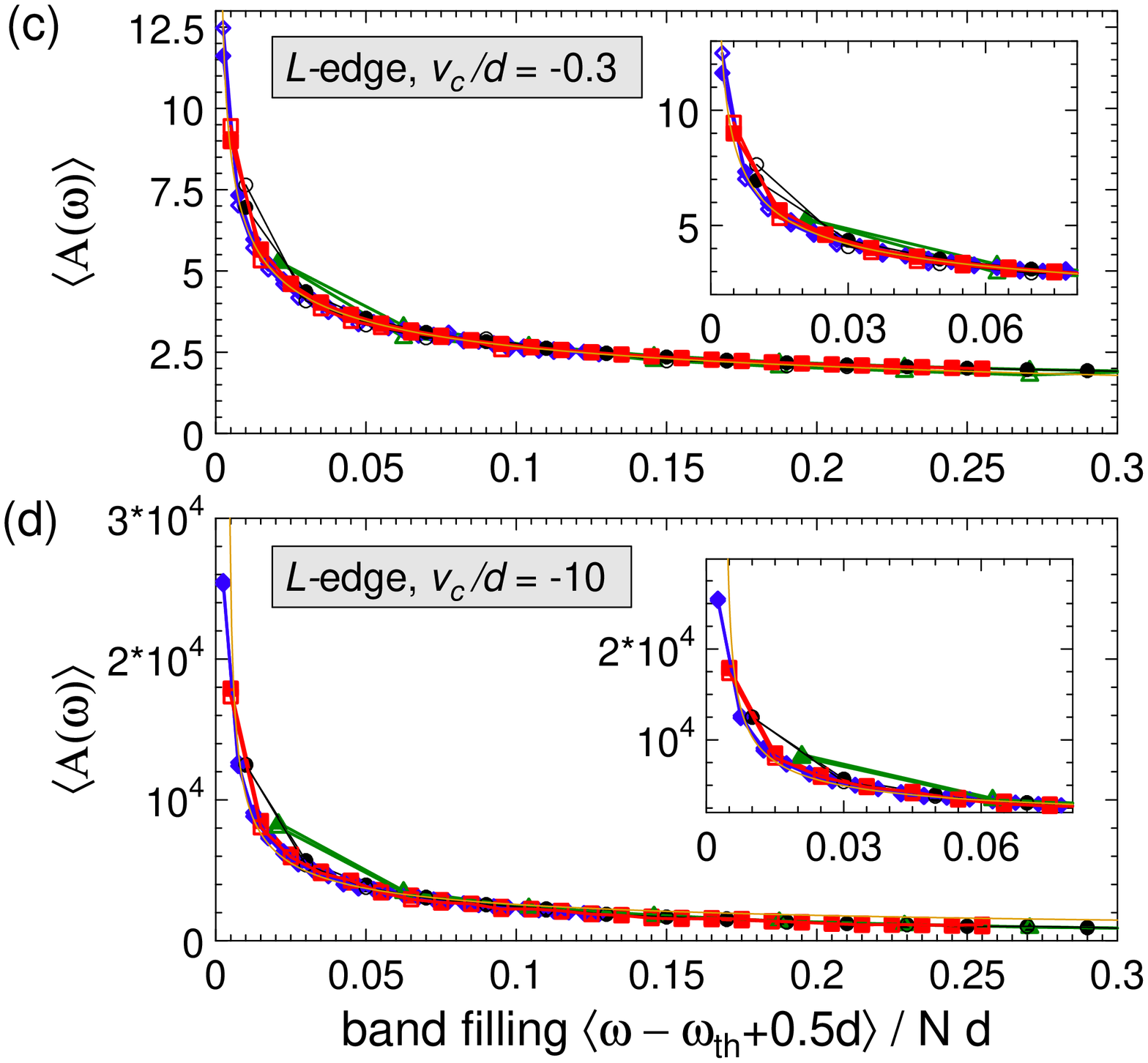}
\caption{(Color online) Mesoscopic averaged photo-absorption cross section as a function of the number $2M$ of particles in a half-filled band for (a)-(b) K-edge and (c)-(d) L-edge, and both weak coupling [(a),(c) $\vc/d \!=\! -0.3$] and strong coupling [(b),(d) $\vc/d \!=\! -10$].
Results are for the COE with full spin up to excitation energies a quarter of the band width above threshold, and are normalized by the bare photo-absorption value.
The K-edge appears, apart from the behavior directly at threshold, rounded, and the rounding increases for more particles in the system. In this sense AOC wins the competition with MND as the thermodynamic limit is approached. However, the slight peak at the edge persists as the signature characteristic of a mesoscopic-coherent system. The L-edge clearly is peaked, and this peak sharpens with increasing particle number. For strong coupling, the L-edge is completely dominated by replacement through the bound state, making the magnitude at the L-edge much larger.
  \label{fig_photoabsvcNdep}}
\end{figure*}

The dipole matrix elements $w_{c \mu}$ for the L-edge are proportional to the wave-function $\psi_{\mu}(\brc) \!=\! \tilde u_{\mu}$ at the position of the perturbation (see Sec.~\ref{sec_dipole}). We therefore find that whenever the perturbation is strong enough to form the bound state this latter will play a very significant role. Indeed, in the situation where the $l\!=\!0$ channel provides all the screening of the impurity (i.e. $\delta_0 \!=\! \pi/2$), the photo-absorption process is {\em entirely dominated} by the term $\omega_{c0} \tilde c^\dagger_0 c_c$ in the dipole operator Eq.~(\ref{eq:dipoleoperator}). The shape of the average photo-absorption spectra can in this case be essentially understood from the energy dependence of the overlap $\Delta_{\bar{0}\gamma}$ shown in Fig.\,\ref{fig_reploverlap}: a sharp peak for $\gamma\!=\!M$ is followed by a relatively long tail. As shown in Figs.\,\ref{fig_Ledge_avgspec}, this is the behavior of the L-edge photo-absorption spectrum in the strong perturbation regime. Note that the L-edge peak is considerably stronger than for the K-edge (compare to Fig.\,\ref{fig_Ledge_avgspec}). Convolution of the sharp peak in the active channel with the spectator spin result still yields a prominent peak. 

In comparing with the bulk-like case, we start by emphasizing again that the physics for an L-edge in the strong perturbation regime is dominated by the bound state. As there are no strong differences between the bound state in the bulk-like and mesoscopic cases, we expect the results to be similar. Fig.\,\ref{fig_KLedge} shows, in fact, a stronger result: for a strong perturbation, the mesoscopic and bulk-like spectra at an L-edge agree quantitatively.\cite{bulkLedge_differs}

\subsection{Dependence on Perturbation Strength and Number of Particles}

So far our major focus has been the strong perturbation regime and a model nanosystem with 100 electrons (50 electrons per spin species in a half-filled band). Since the processes that determine the shape of the edge, namely the AOC and MND responses, depend on the number of particles, we will now address how the (average) photo-absorption spectra change as the number of particles in the system is varied. This is of particular interest with respect to experiments since the number of electrons in the system can be adjusted by means of the geometry (size), gate voltages, or the density of states (doping).
It is convenient at the same time to vary the strength of the perturbation produced by the core hole. The strong perturbation regime describes semiconductor heterostructures. In these systems, only $s$-conduction electrons are present and available for screening; the Friedel sum rule then implies the strong perturbation regime. In contrast, Fermi edge singularities in metals are described by weaker perturbations,\cite{tanabe:RMP1990, citrin:PRB1979} corresponding to a smaller phase shift for the $s$-electrons and in agreement with the fact that other channels of conduction electrons ($p$, $d,...$) are involved in the screening. Metallic nanoparticles are one mesoscopic system where similar values for the phase shifts occur. Other situations where the small perturbation regime is relevant could arise.

Fig.~\ref{fig_photoabsvcNdep} shows photo-absorption spectra at both the K- and L-edge for two different perturbation strengths with the number of electrons ranging from $24$ to $200$. (Note that COE statistics are used here rather than the CUE statistics used in Figs.~\ref{fig_Kedge_avgspec}-\ref{fig_Ledge_avgspec}.) The weaker perturbation ($\vc\!=\!-0.3$) produces a phase shift at $E_F$ which is typical of a metallic environment, while the larger strength produces complete $s$-wave screening ($\delta_{0F}\!\approx\!-\pi/2$) suitable for semiconductors. All curves show the main FES signatures: for a K-edge, a peak at threshold superposed on a rounded edge, while at an L-edge, a strong peak at threshold. Clearly, the FES signatures are enhanced when the perturbation is strong.

For the weaker perturbation, the bound state is not formed. Absorption near the edge is nevertheless enhanced because of correlation between the spectrum and the values of the wave-functions at $r_c$. This implies in particular that all the terms in the replacement sum Eq.~(\ref{eq:dirreplthres}) have the same sign, as in the bulk and despite the random character of the wave-functions.


Since the mean level spacing $d\!=\!2 \pi/N$ is $N$-dependent, the energy from threshold is given here in units of band filling above threshold, $\la \omega \!-\! \omth \ra\,/\,N\, d$ (such that an excitation into the highest level $\gamma \!=\! N\!-\!1$ corresponds to a value $1/2$) in order to allow for a direct comparison of the curves. Note that we do not confine our attention to the immediate threshold vicinity but rather consider excitation energies that allow electrons to fill states up to $3/4$ of the band width.

We first discuss the K-edge spectra in Fig.~\ref{fig_photoabsvcNdep}. First, note that all four curves converge at large energy to the same behavior. In fact, they approach a line corresponding to the bare process ($\sim\! |w_{c\gamma}|^2$). This is easily understood: Once the photon energy is well above threshold, so that $\omega \!-\!\omth$ is significantly larger than the width of the unperturbed ground state in the perturbed basis (see Fig.~\ref{fig_GSwidth}), the final states are necessarily of the form 
$ | \Psi_f \ra \!=\! \tilde{c}^\dagger_\kappa | \Psi^0_f \ra$ where $| \Psi^0_f \ra$ is one of the states with a significant overlap with $\Phi_0$. All the ``extra energy'' is borne by the highly excited one particle state $\kappa$. In this case, only the term $\omega_{c\kappa} \tilde{c}^\dagger_\kappa c_c$ of the dipole operator $\hat D$ contributes to the absorption -- there are no replacement processes. Summing over final states amounts to averaging $\omega_{c\kappa}^2$ on an energy window equal to the width of $\Phi_0$ in the perturbed basis, and so the result is the same as for the bare process.

Moving toward the K-edge threshold, one sees that the photo-absorption is first suppressed and then jumps right at threshold. The suppression is simply the manifestation of the AOC familiar from the bulk K-edge. In fact, as $N$ increases, the rounding becomes more pronounced. When the perturbation is weak, the resulting points lie right on the bulk power-law curve. Thus, for weak perturbation, the average mesoscopic photo-absorption in the $N \!\to\! \infty$ limit yields the bulk singularity. For strong perturbation (the semiconductor case), there are substantial deviations from the bulk power-law: at large energies the average mesoscopic photo-absorption is suppressed while for energies just above threshold, as for the threshold point itself, photo-absorption is enhanced.

A striking difference between the bulk and mesoscopic K-edge spectra occurs right at threshold: rather than following a rounded edge, the average mesoscopic response shows a peak (see also Fig.~\ref{fig_KLedge}). The magnitude of this peak does depend on the strength of the perturbation but is approximately $N$-independent. For the strong perturbation case ($\vc \!=\! -10$), the peak is approximately $50\%$ larger than the second point and about $3$ times larger than the bulk result. This enhancement near the Fermi edge comes from the dipole matrix elements which, for the generic nanosystems considered here, are non-zero even at the K-edge: breaking of rotational symmetry allows MND processes in the mesoscopic case while only AOC is present in bulk.

The L-edge spectra in Fig.~\ref{fig_photoabsvcNdep} show the strongly peaked edge characteristic of the MND singularity. In the case of a weak perturbation, panel (c), there is extremely good agreement between the average mesoscopic photo-absorption and the bulk power-law response for all energies and all $N$. Right at threshold, it appears that the average mesoscopic result is slightly larger than the bulk-like result. The peak becomes more pronounced as $N$ increases because, as for the K-edge, one is able to access smaller energies with respect to the band width. 

For strong perturbation at the L-edge, note the very large magnitude of the photo-absorption. This stems from the fact that replacement processes through the bound state completely dominate. In this case, the bulk power law provides a qualitative guide to the peak, but, as for the K-edge, does not agree quantitatively.




\section{Fluctuations of the Photo-Absorption Cross Section}
\label{sec_results_fluct}

\begin{figure}[b]
\includegraphics[width=8.5cm]{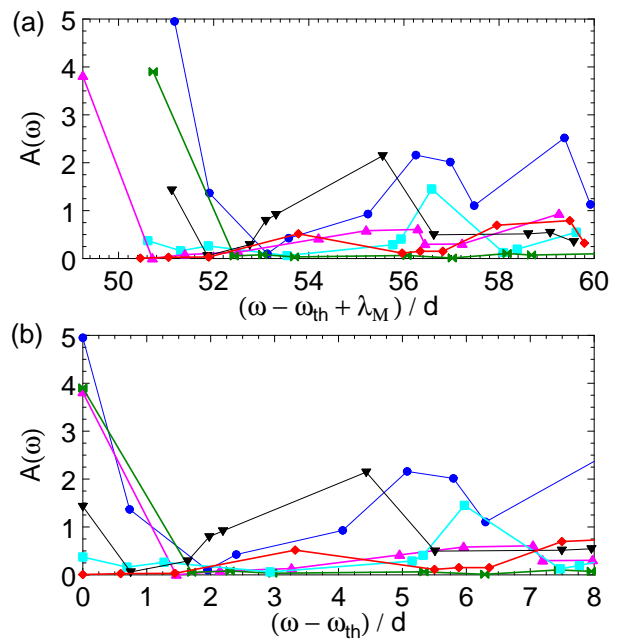}
\caption{(Color online) Individual K-edge absorption spectra of six mesoscopic samples (marked by different symbols), illustrating the outcome expected in real single-sample measurements ($N\!=\!100, M\!=\!50, \vc/d\!=\!-10$, COE). Fluctuations occur in both the energy and the cross section. The spectra consist, strictly speaking, of a series of $\delta$-functions (experimentally broadened) whose weight and positions are marked by the points; the lines are guides for the eye. (a) Photo-absorption cross-section as a function of photon energy $\omega$. The cross section is highly fluctuating, not necessarily peaked, and can even zigzag. Both the threshold energy and the distance between the points vary on the scale of the mean level spacing. (b) Spectra shifted such that their threshold energies coincide. A peaked edge visibly emerges.
  \label{fig_indivspec}}
\end{figure}

\begin{figure}
\includegraphics[width=8.5cm]{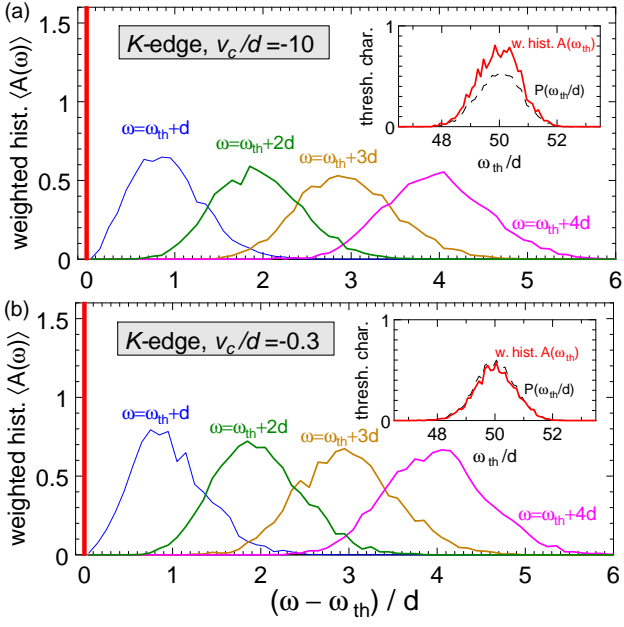}
\caption{(Color online) Fluctuations of the photo-absorption cross section at the K-edge for (a) strong and (b) weak perturbation ($N\!=\!100$, $M\!=\!50$, CUE, optically active spin). The average
photo-absorption cross section as a function of energy, $\langle A(\omega) \rangle$, is shown for processes with different average excitation excitation (marked $\omth\!+\!d$,  $\omth\!+\!2d, \ldots$). Note that energies are measured with respect to the threshold energy $\omth$ (vertical line) as would be the case in experiments. The behavior of the threshold energy and cross section are shown in the inset: both are approximately Gaussian distributed. The peak next to threshold is clearly asymmetric with the maximum photo-absorption cross section found at energies distinctly below the average value. The curves broaden and symmetrize away from threshold. The area under the curves is the total photo-absorption; the (slight) peak at threshold is seen by noticing that the area in the inset is largest, as well as being greater for larger perturbations.
\label{fig_fluctK}}
\end{figure}

\begin{figure}
\includegraphics[width=8.5cm]{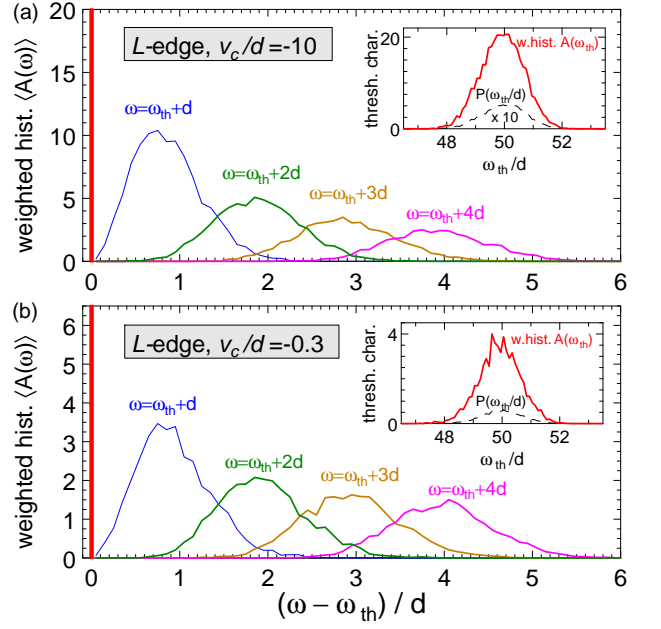}
\caption{(Color online)
Fluctuations of the photo-absorption cross section at the L-edge for (a) strong and (b) weak perturbation ($N\!=\!100$, $M\!=\!50$, CUE, optically active spin). Explanations are the same as for Fig.~\ref{fig_fluctK}. The strongly peaked FES is evident.
\label{fig_fluctL}}
\end{figure}

\begin{figure}
\includegraphics[width=8.5cm]{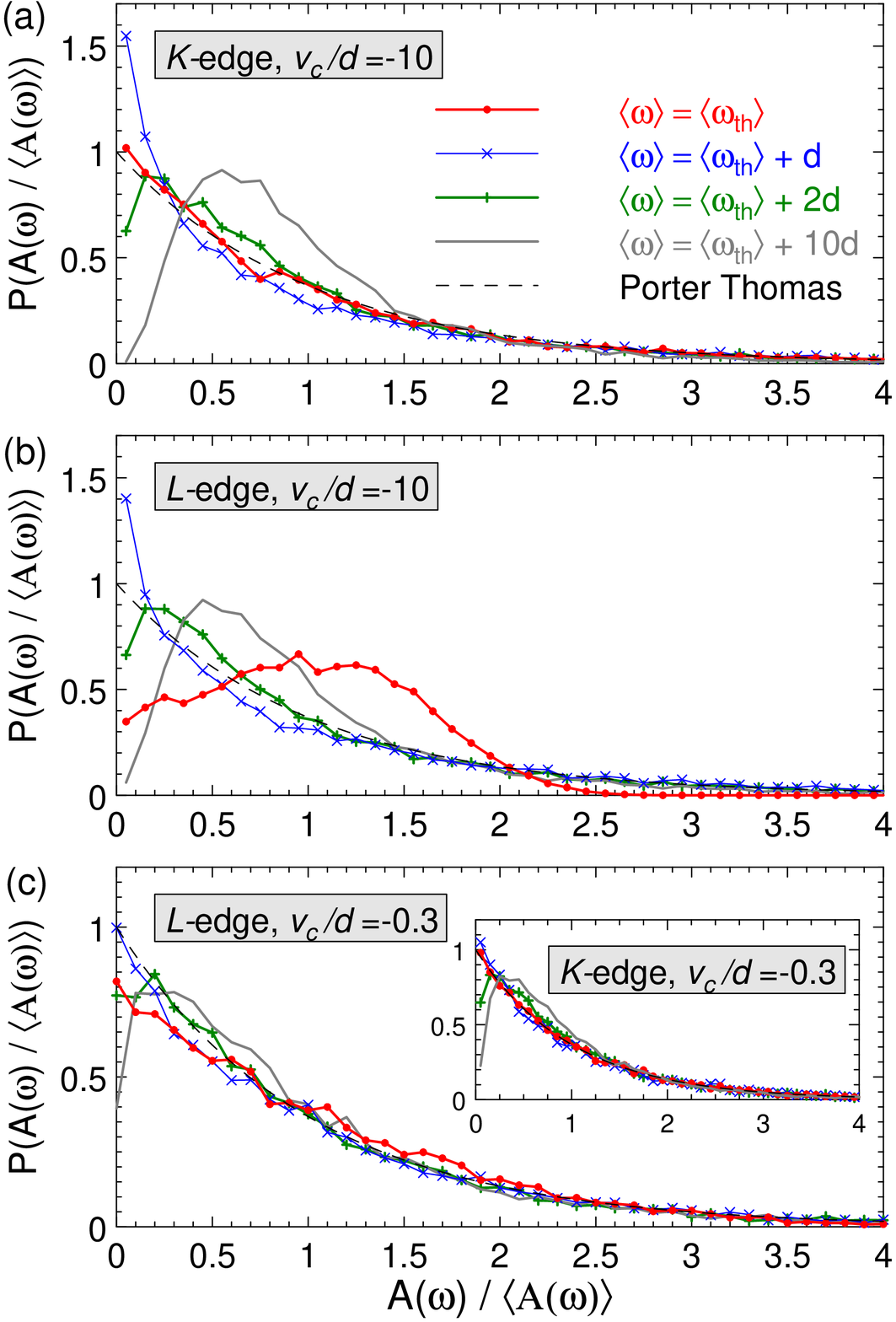}
\caption{(Color online) Distribution of the mesoscopic photo-absorption fluctuations ($N\!=\!100$, $M\!=\!50$, CUE): for a strong perturbation $\vc / \Deltaml \!=\! -10$, (a) the K-edge and (b) the L-edge, and for a weak perturbation $\vc / \Deltaml \!=\! -0.3$ at (c) the L-edge with the K-edge in the inset. The probability distribution of the photo-absorption $A(\omega)$ normalized by its mean value $\la A(\omega) \ra$ is shown for different mean excitation energies $\la \omega \ra$, both near threshold (curves with symbols) and further away from threshold (full line) where a Gaussian shape emerges. For comparison, a Porter-Thomas distribution is indicated by the dashed line. Near threshold, the distributions are Porter-Thomas in almost all cases because the fluctuations in the dipole matrix elements dominate and cancel all correlations from the overlap. An exception occurs at threshold for the L-edge where the excited electron sits in the first level above the Fermi energy. In this case, the distribution resembles that of a ground state overlap.
\label{fig_photoabsfluct}}
\end{figure}

Turning from the average photo-absorption, we now investigate a quantity inherent to mesoscopic systems and a key characteristic of the photo-absorption cross section: its fluctuations. Fig.~\ref{fig_indivspec} shows several examples of the photo-absorption cross-section of individual systems. As for the average, all data presented are for absorption in the primary band and are normalized to the total absorption of the primary band. There are two sources of fluctuations: wave-function (matrix-element) fluctuations and energy-level fluctuations.

First, the wave-function amplitudes at the location of the core hole vary, causing the dipole matrix elements to fluctuate. This is the most dramatic mesoscopic effect, causing large fluctuations in the photo-absorption cross section (Fig.~\ref{fig_indivspec}). In particular, the shape of the spectrum of an individual system is not necessarily peaked, but can be ``rounded'', almost uniform, or even zigzag.

Second, the energy level effect creates fluctuations in the photo-absorption as well as in the photon energies that can be absorbed.
The former is small compared to the effect of the wave-functions, but the latter can be quite significant. The absorbed photon energies fluctuate by an amount of order the a mean level spacing $\Deltaml$, with increasing width further away from threshold. Because core electrons at different locations can be excited by the different photons, subsequent measurements of one and the same system will result in different final (perturbed) energy levels $\{\lam\}$, and therefore give rise to different spectra even though the unperturbed energy levels $\{\eps\}$ remain the same. With single measurements of different systems, as we assume here, both the $\{\eps\}$ and $\{\lam\}$ would vary.

Note that the threshold energy varies from dot to dot, resulting in a relative shift of the spectra of a few mean level spacings [Fig.~\ref{fig_indivspec}(a)]. In Fig.~\ref{fig_indivspec}(b) the spectra are shown with respect to their threshold energy, clearly enhancing the organization of the ``measurement'' points. Still, the fluctuations add up away from threshold such that the spectra can span rather different energy ranges.

In order to analyze the fluctuations more systematically, Figs.~\ref{fig_fluctK} and \ref{fig_fluctL} show the photo-absorption cross section as a weighted histogram, highlighting the fluctuations in energy weighted by absorption intensity. The total cross section is broken down by grouping processes which have the same mean photon energy, and then for each of these groups the average $\langle A(\omega) \rangle$ is plotted. Two perturbation strengths are contrasted, and energies are measured with respect to the threshold energy (as would be the case in experiments). Note that the curves for the different classes have substantial overlap. For the peak next to threshold, higher cross sections are achieved at smaller energies for all perturbation strengths and at both the K- and L-edges. Further away from the Fermi edge, the histograms become symmetric, broader, and more Gaussian, as one expects from the Central Limit Theorem.

In the rest of this Section, we focus on the statistical properties of the absorption amplitude. We aggregate all the processes (final states) taking place at the same mean energy; the resulting probability distributions are shown in Fig.~\ref{fig_photoabsfluct}.

Most of the curves closely resemble the Porter-Thomas distribution, in particular for the weaker perturbation (exceptions are discussed below).  This originates from the dipole matrix elements $\omega_{c\kappa}$ being proportional to Porter-Thomas distributed wave-function derivatives (K-edge) or wave-functions (L-edge) at the position of the perturbation (Sec.~\ref{sec_dipole}). The corresponding randomness overwhelms any correlations in the ground state, replacement, or shake-up overlaps. Further away from threshold, as more and more processes are included, the distribution becomes more Gaussian.

At the L-edge in the limit of strong perturbations, we find however clear
deviations from the Porter-Thomas distribution
[Fig.~\ref{fig_photoabsfluct}(b)]. At threshold, $\la \omega \ra \!=\!
\la \omth \ra$, the corresponding curve (circles) is very similar to
the probability distribution $P(|\Deltaov|^2)$ of the ground state
overlap (between the unperturbed and perturbed many-body ground state)
presented in the first paper of this series (see Fig.~3
of Ref.\,\onlinecite{xrayaoc}). The origin of this behavior can be
traced back to the fact that, as discussed in
Sec.~\ref{sec_boundstate}, the absorption in the L-edge strong perturbation
regime is entirely dominated by processes involving the bound state.
Therefore at threshold
\begin{equation}
      A(\omth) \simeq |w_{c0}|^2 |\Deltaov_{\bar{0} \, M}|^2 \; .
\end{equation}
As $|w_{c0}|^2$ does not fluctuate, the distribution of $A(\omth)$ is the same as that of $|\Deltaov_{\bar{0} \, M}|^2$, which is equivalent\cite{xrayaoc} to an Anderson ground state overlap for positive perturbation and phase shift $\pi \!-\! |\delta_F|$ [Eq.~(21) in Ref.\,\onlinecite{xrayaoc}]. In the strong perturbation limit, $|\delta_F|\!=\!\pi/2 = \pi \!-\! |\delta_F|$, and therefore the distribution of $|\Deltaov_{\bar{0} \, M}|^2$ is {\em the same} as the ground state overlap distribution. For weaker perturbations, such that the bound state still dominates the absorption at threshold but with a phase shift $|\delta_F| \!<\! \pi/2$, the symmetry relation between the Anderson overlap for negative perturbation and the replacement overlap involving the bound state for a positive perturbation still holds, but the negative and positive phase are no longer equal.

The mechanism underlying the fluctuations in the L-edge strong perturbation case when the photon energy is above threshold deserve further discussion.  On the one hand, as illustrated in Fig.~\ref{fig_GSwidth}, the overlap between final states $\Psi_f \!=\! c^\dagger_0 \Psi^0_f$ and the unperturbed ground state rapidly becomes extremely small. On the other hand, the bound state dipole matrix element $w_{c0}$ is much larger than the others (basically $|w_{c0}|^2$ is of the same order of magnitude as $\sum_{i \neq 0} |w_{ci}|^2 $).  It is therefore not clear a priori which effect, the smallness of the overlaps or the largeness of the matrix element, will dominate.

\begin{figure}
\includegraphics[width=8.5cm]{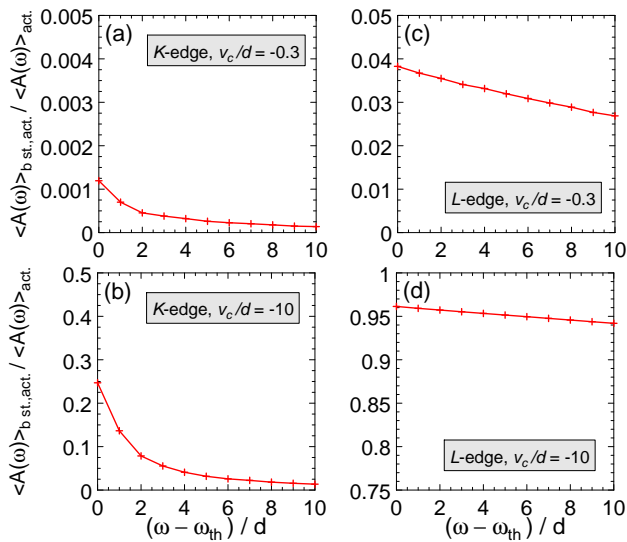}
\caption{(Color online) Contribution of bound state processes to the average photo-absorption cross section as a function of photon energy above threshold ($N\!=\!100$, $M\!=\!50$, CUE, optically active spin). All replacement and shake-up processes involving the bound state are included. Upper (lower) panels: weak (strong) perturbation. Left panels: K-edge. Right panels: L-edge. For weak perturbation, the contribution is, as expected, small. For strong perturbation, processes through the bound state make a significant but decaying contribution at the K-edge; at the L-edge, they dominate.
\label{fig_ImportanceOfBS}}
\end{figure}

The answer to this question is given in Fig.~\ref{fig_ImportanceOfBS}, which shows the relative importance of bound state dominated processes in various cases. We see, first, that for weak perturbations, the importance of the lowest perturbed one-particle state (which is not here properly speaking a bound state) is, as expected, marginal. We further see that for the K-edge with strong perturbation, this contribution is not negligible (about $25\%$ near the edge for the case considered) but neither is it dominant. Furthermore, it decays rapidly to zero as the photon energy becomes larger than the width of the unperturbed ground state in the perturbed basis.

For the strong perturbation L-edge case, panel (d), which is our main interest here, we see that the largeness of the dipole matrix element is the dominant effect, even at large photon energy. The overlap does not decrease fast enough to compensate for this, and only a moderate decrease of the relative contribution of the boundstate is observed. As a consequence, there are no fluctuations associated with the dipole matrix element (since $|w_{c0}|$ does not fluctuate)--- the fluctuations derive entirely from those of the overlap  between the unperturbed ground state $|\Phi_0 \rangle$ and the final states $|\Psi_f \rangle$ from which the bound state has been removed. As the energy of the photon increases, the number of shake up processes (i.e.\ of final states) increases, and the fluctuations become more Gaussian.

So far we have discussed fluctuations in the CUE situation where time-reversal symmetry is broken, corresponding to the average photo-absorption spectra shown in Sec.~\ref{sec_results_avg}. Very similar results apply to the COE case (presence of time-reversal symmetry) as we found above for the average photo-absorption. The major difference comes from the change in the Porter-Thomas distribution (it now diverges at zero); once this is taken into account, the fluctuation characteristics (to the extent described here) are the same for COE and CUE.

\section{Experimental Realizations}
\label{sec_exp}

We end this paper by discussing some possible experimental realizations of the physics we have described. Beyond a simple illustrative purpose, this Section provides an opportunity to discuss in more detail the applicability of the model we have used to real physical systems.

The model in Eqs.~(\ref{eq:hamH0})-(\ref{eq:ham}) has been built to contain the main ingredient of the physics under consideration, namely the sudden appearance of a local perturbation whose phase shift {\em at the Fermi energy} is determined by the Friedel sum rule. It turns out, however, that the output of the model does not depend only on its properties at the Fermi energy, even qualitatively.  In particular, the bound state created in the strong perturbation regime plays a major role in influencing both the mean and fluctuations of the photo-absorption spectrum. It is therefore useful to discuss for a few examples how much this mechanism is expected to be taken into account correctly.

Since  the  basic physics  we  describe -- a localized  (rank  one) perturbation acting on a Fermi sea of chaotic electrons -- is very general, it can be realized in various systems:

(1) An obvious experiment would be the direct realization of x-ray absorption measurements in small metallic nanoparticles. Either colloidal nanoparticles with diameter 1-2 nm or ``metallic molecules" such as Au$_{55}$ might be used. The technical requirements for such an experiment are beyond standard capability at present: Although the spatial resolution of x-ray micro-beams has been demonstrated at low temperature, their energy resolution seems to be insufficient \cite{xraymicrobeam}. Such an experiment may, of course, be possible in the future.

With regard to the importance of the bound state (discussion at the end of Section \ref{sec_boundstate}), the bound state should play the same general role in a nanoparticle as in the bulk material. In the bulk metals that have been studied most carefully \cite{tanabe:RMP1990,citrin:PRB1979}, several angular momentum channels are involved in screening, yielding phase shifts less than 1. These, then, are in the weak perturbation regime: the bound state plays a minor role in both the bulk and mesoscopic situations.

(2) Another, intrinsically different, example occurs in a double quantum dot where the tunneling of an electron into (or out of) the system causes a rank one perturbation due to the constriction which mediates a sudden change of the wave-function in one dot to the other \cite{matveev,levitov}. The AOC analysis in Ref.\,\,\onlinecite{matveev} shows that the phase shifts involved are $\pm \pi/4$. Such a phase shift is produced by $\vc / \Deltaml$ in the range $-0.3$ to $-0.5$. Thus for this situation as well, the perturbation is weak, and the effect of the bound state is minor.

(3) Finally, photo-absorption in quantum dots is a third example, one which is feasible with existing standard semiconductor heterostructure technology \cite{dieterweiss}. Since for semiconductor conduction electrons all of the screening comes from a single channel, the phase shift is $\pi/2$, and the role of the bound state is critical. We now  describe in detail this possibility.

\begin{figure}[t]
\includegraphics[width=7.8cm]{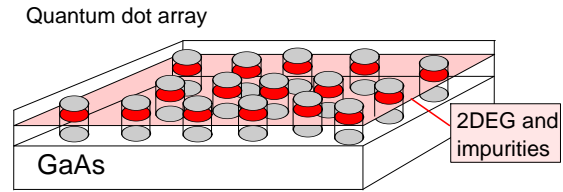}
\caption{(Color online) Experimental arrangement to test our prediction of a transition from a rounded K-edge to a slightly peaked edge as the system size is diminished to the mesoscopic coherent regime. We propose a quantum dot array in a semiconductor heterostructure where the plane of the 2DEG also contains impurities that provide a localized state in the GaAs band gap.
  \label{fig_expreal}}
\end{figure}

First, a quantum dot array can be formed in a semiconductor heterostructure like GaAs/AlGaAs using standard growth and etching techniques (see Fig.~\ref{fig_expreal}). The dot diameter can be made smaller than 100 nanometers, with a variation of $\pm 5$ nanometers, using state-of-the-art techniques.  This gives a mean level spacing of about $30\!\pm\!3$ $\mu$eV. Slight fabrication imperfections guarantee that the dots are chaotic, all slightly different but with a well defined average. The layer of the two-dimensional electron gas (2DEG) cuts through all the dots which ``stand'' as pillars on a GaAs substrate. Each dot contains several tens of conduction electrons; this number can be varied in the fabrication process by adjusting the doping.

The core electron of the original x-ray edge problem is now replaced by an electron in an impurity level within the GaAs bandgap. In order to realize the assumptions made in the paper, the electron has to be in a localized state, and in order to probe a K-edge situation, it has to have $s$-like symmetry. This can be realized using a suitable deep level, or by doping with N (isoelectronic to As). The impurities host a $s$-type bound state occupied by an electron that is subsequently excited into the conduction band. The variety of available dopants should even allow one to change the symmetry of the impurity level, and thus to study the L-edge behavior. The impurities have to be placed into the 2DEG in order to ensure a sufficient overlap and have a suitable density (few impurities per dot).

The excitation of the impurity-pinned electrons is achieved using a micrometer laser (rather than an x-ray). The resolution of these lasers is around $10$ $\mu$eV, thus below the mean level spacing of the electrons in the quantum dots and sufficient to resolve the expected effects within a few mean level spacings from threshold. The photo-absorption cross section has to be measured at low temperature.

The first quantity investigated in this paper is the average photo-absorption cross section. Although it is tempting to measure it directly as the photo-absorption of the array, this may not be possible (yet) due to fabrication uncertainty. The Fermi energy in each dot (measured from the impurity level) will vary by about a mean level spacing from dot to dot due to single particle effects; in addition, the charging energy may lead to a slightly larger variation, perhaps two mean level spacings. Finally, the mean level spacing itself will vary by about 10\% from dot to dot (assuming 5\% uncertainty in the linear dimension). Taken together, the dot-to-dot variation may wash-out the expected threshold peak in the photo-absorption, which extends, after all, over only about two mean level spacings. Therefore, each dot may have to be measured individually. Afterwards, the average over an ensemble (the array) has to be performed.
In this way, information about the fluctuations can, of course, be extracted as well.

\section{Conclusions}
\label{sec_summary}

In this paper we have studied Fermi edge singularities in the mesoscopic regime, in particular photo-absorption spectra of generic, chaotic-coherent mesoscopic systems with a finite number
of electrons. 
The basic underlying physics that we study is, however, much more general, and the model can be applied to any situation where a sudden, localized (rank one) perturbation acts on a finite number of chaotic electrons via a (dipole) matrix element.

The photo-absorption cross section differs from the naive expectation due to two counteracting many-body responses resulting from the sudden perturbation of the system when a core or impurity electron is removed: the Anderson orthogonality catastrophe (AOC) and the Mahan-Nozi\`{e}res-DeDominicis (MND) contribution. We have studied AOC in detail in the first paper of this series \cite{xrayaoc}. We found, for instance, that typical mesoscopic systems are rather far from the thermodynamic (bulk) limit and AOC is incomplete.
This is also reflected in the photo-absorption spectra: the K-edge photo-absorption cross section near threshold remains finite.

Whereas the threshold behavior in x-ray photoemission (excited electron leaves the system) can be deduced from the power law exponent of the AOC overlap alone, this does not apply to photo-absorption or luminescence experiments (where the electron is excited into, or relaxes from, the conduction band) where, besides AOC, the MND response is crucial. It counteracts the AOC effect if the dipole selection rules are fulfilled. In the present paper we included this response and computed photo-absorption spectra of generic mesoscopic systems in various situations using a Fermi golden rule approach. In metals the result is typically a rounded K- and a peaked L-edge in the photo-absorption cross section.

We confirm the peaked L-edge behavior in the average photo-absorption cross section for mesoscopic systems. At the K-edge, however, we find characteristic changes that allow one to infer the presence of chaotic-coherent dynamics of electrons in nanosystems from the near-threshold behavior of the average photo-absorption. As the system size is diminished to reach the mesoscopic-coherent scale, the most important change occurs in the dipole matrix elements. The chaotic dynamics of the (conduction) electrons makes them distinctively different from the bulk-like situation: The electronic wave-function and its derivative at a certain position are independent, and the dipole matrix elements at the K-edge ($\propto \psi'_{\kappa}$) are non-zero. As a consequence, mesoscopic K-edge spectra are peaked and not rounded as in the metallic (bulk-like) case. This peak is visible right at threshold and is a direct signature of chaotic-coherent dynamics of electrons in mesoscopic systems. We propose in detail experiments where this prediction can be tested using current semiconductor technology.

The fluctuations of the photo-absorption cross section are largely
dominated by the Porter-Thomas distribution of the dipole matrix
elements. However, deviations occur at the L-edge right at threshold
when the cross section probability distribution resembles that of
the AOC ground state overlap. The reason for this behavior is the dominance of {\it replacement through the bound state}, a recurring theme in this subject.


\begin{acknowledgments}
M.H. gratefully thanks Prof.~Ohtaka for illuminating discussions at Chiba University that clarified the relation of this study to early work on the x-ray edge problem. We thank K.~Matveev for several valuable discussions, and I.~Aleiner, Y.~Gefen, T.~Harayama, I.~Lerner, E.~Mucciolo, Y.~Ochiai, I.~Peschel, I.~Smolyarenko, Y.~Tanabe, U.~R\"o{\ss}ler, D.~Weiss, and W.~Wegscheider for useful conversations. M.H. acknowledges the hospitality of ATR (Kyoto, Japan) and the support of both the Alexander von Humboldt Foundation and the Deutsche Forschungsgemeinschaft (DFG) through the Emmy-Noether Program. The work in the U.S.\ was supported in part by the NSF (DMR-0506953).
\end{acknowledgments}



\begin{thebibliography}{99}

\bibitem{mahan:book} G.~D.~Mahan, \emph{Many-Particle Physics}, 3rd edition,
Kluwer Academic/Plenum Publishers, New York, 2000.

\bibitem{nozieres} B.~Roulet, J.~Gavoret, and P.~Nozi\`{e}res,
Phys.~Rev. {\bf 178}, 1072 (1969);
P.~Nozi\`{e}res,  J.~Gavoret, and B.~Roulet,
Phys.~Rev. {\bf 178}, 1084 (1969);
P.~Nozi\`{e}res and C.~T.~De Dominicis,
Phys.~Rev. {\bf 178}, 1097 (1969).

\bibitem{tanabe:RMP1990}  For  a review  and  further references,  see
Y.~Tanabe and O.~Ohtaka, Rev.~Mod.~Phys.~{\bf 62}, 929 (1990).

\bibitem{xrayprl} M.~Hentschel, D.~Ullmo, and H.~U.~Baranger,
Phys.~Rev.~Lett. {\bf 93}, 176807 (2004) [arXiv:cond-mat/0402207].

\bibitem{xrayaoc} M.~Hentschel, D.~Ullmo, and H.~U.~Baranger,
Phys.~Rev. B {\bf 72}, 035310 (2005) [arXiv:cond-mat/0503330].

\bibitem{anderson:PRL1967}
P.~W.~Anderson, Phys.~Rev.~Lett.~{\bf 18}, 1049 (1967).

\bibitem{kondo_hewson} A.~C.~Hewson, {\it The Kondo Problem to Heavy Fermions},
Cambridge University Press, Cambridge, UK (1993).

\bibitem{schotteschotte} K.~D.~Schotte and U.~Schotte, Phys.~Rev. {\bf 182}, 479 (1969).

\bibitem{friedelscomment} J.~Friedel, Comments Solid State Phys. {\bf 2}, 21 (1969).

\bibitem{hopfield} J.~J.~Hopfield, Comments Solid State Phys. {\bf 2}, 40 (1969).

\bibitem{tanabe:seriesofpapers} 
Y.~Tanabe and K.~Ohtaka, Phys. Rev. B {\bf 28}, 6833 (1983);
Phys. Rev. B {\bf 29}, 2036 (1984);
K.~Ohtaka and Y.~Tanabe, Phys. Rev. B {\bf 30}, 4235 (1984);
Phys. Rev. B {\bf 34}, 3717 (1986);
Phys. Rev. B {\bf 39}, 3054 (1989).

\bibitem{citrin:PRB1979}
P.~H.~Citrin, G.~K.~Wertheim, and M.~Schl\"uter, Phys.~Rev.~B {\bf 20}, 3067
(1979).

\bibitem{ohtaka:private} K.~Ohtaka, private communication.

\bibitem{vallejos:PRB2002} R.~O.~Vallejos, C.~H.~Lewenkopf, and
Y.~Gefen, Phys.~Rev.~B {\bf 65}, 085309 (2002).

\bibitem{gefen:PRBR2002}
Y.~Gefen, R.~Berkovits, I.~V.~Lerner, and B.~L.~Altshuler,
Phys.~Rev.~B {\bf 65}, 081106(R) (2002).

\bibitem{kroha:PRB1992} Y.~Chen and J.~Kroha, Phys.~Rev.~B {\bf 46}, 1332 (1992).

\bibitem{calleja}
J.~M.~Calleja, A.~R.~Go\~{n}i, B.~S.~Dennis, J.~S.~Weiner, A.~Pinczuk, S.~Schmitt-Rink,
L.~N.~Pfeiffer, K.~W.~West, J.~F.~M\"uller, and A.~E.~Ruckenstein,
Sol.~St.~Comm. {\bf 79}, 911 (1991).

\bibitem{oreg:PRB1996} Y.~Oreg and A.~M.~Finkelstein,
Phys.~Rev.~B {\bf 53}, 10928 (1996).

\bibitem{mehta} M.~L.~Mehta, \emph{Random Matrices}, Academic Press Inc., San Diego, 1991.

\bibitem{bohigas} O.~Bohigas, in {\it Chaos and Quantum Physics}, edited by M.-J.~Giovanni,
A.~Voros, and J.~Zinn-Justin (North-Holland, Amsterdam, 1991), pp.~87-199.

\bibitem{UllmoBaranger01}
D. Ullmo and H. U. Baranger, Phys. Rev. B \textbf{64}, 245324 (2001) [arXiv:cond-mat/0103098]; G. Usaj and H. U. Baranger, Phys. Rev. B \textbf{66}, 155333 (2002) [arXiv:cond-mat/0203074].


\bibitem{smolyarenko:PRL2002}
I.~E.~Smolyarenko, F.~M.~Marchetti, and B.~D.~Simons,
Phys.~Rev.~Lett.~{\bf 88}, 256808 (2002).

\bibitem{matveev:PRL1998}
I.~L.~Aleiner and K.~A.~Matveev, Phys.~Rev.~Lett.~{\bf 80}, 814 (1998).

\bibitem{prigodin}
V.~N.~Prigodin, N.~Taniguchi, A.~Kudrolli, V.~Kidambi, and S.~Sridhar,
Phys.~Rev.~Lett.~{\bf 75}, 2392 (1995).

\bibitem{friedel_boundstate}
J.~Friedel, Phil. Mag. {\bf 43}, 153 (1952). 

\bibitem{combescotnozieres}
M.~Combescot and P.~Nozi\`{e}res, J.~Phys.~{\bf 32}, 913 (1971). 

\bibitem{zagoskin}
A.~M.~Zagoskin and I.~Aflleck, J.~Phys.~A {\bf 30}, 5743 (1997).

\bibitem{berry_planewave}
M.~V.~Berry, J.~Phys.~A: Math.~Gen. {\bf 10}, 2083 (1977). 

\bibitem{voros_planewave}
A.~Voros, Ann.~Inst.~H.~Poincar{\'{e}} {\bf A24}, 31 (1976); {\bf A26}, 343 (1977). 

\bibitem{metropolis}
We wish to produce sets of eigenvalues $\{\eps\}$ and $\{\lambda\}$ that follow the joint distribution obtained explicitly by Aleiner and Matveev \cite{matveev:PRL1998}. To this end we place all energy levels on the (unit) circle (thereby working with circular, rather than Gaussian, ensembles of random matrices) starting with an equidistant distribution. After a transition time of $\propto 50 N$ steps we start to generate our ``samples'' along the lines of the Metropolis algorithm. To avoid correlations between realization, it has proven useful, besides waiting a sufficient number of steps ( $\propto 4 N$) before a new realization is taken, to move a pair $\{\eps_i, \lam_i\}$ (rather than individual energy levels) in every third step of the algorithm.  If a bound state was formed, it was placed by hand at the respective energy when the Metropolis algorithm was interrupted (i.e., a sample was chosen), and its fluctuations are negligible.  The curves shown are computed in this way typically using 10,000 samples (i.e., uncorrelated realizations of the random energy levels $\{ \eps \}$ and $\{ \lam \}$ with the correct statistics).

\bibitem{onepairshup_dominate} J.~D.~Dow and C.~P.~Flynn,
J. Phys. C: Solid St. Phys., {\bf 13}, 1341 (1980).

\bibitem{bulkLedge_differs}
We note that the definition of the bulk L-edge photo-absorption used here is slightly different from that in our previous work Ref.\,\onlinecite{xrayprl}. There, we took the dipole matrix element to be the same for each level in the perturbed Fermi sea, \textit{including the bound state}. Here, we take the magnitude of the unperturbed wave functions to be the same, and solve for the resulting dipole matrix element to the bound state, which will be large in the strong perturbation limit. 

\bibitem{xraymicrobeam}
M.~Yoon {\it et al.}, Appl.~Phys.~Lett.~{\bf 75}, 2791 (1999);
L.~Kipp {\it et al.}, Nature {\bf 414}, 184 (2001);
C.~G.~Schroer {\it et al.}, Appl.~Phys.~Lett. {\bf 82}, 1485 (2003).

\bibitem{matveev} 
K.~A.~Matveev, L.~I.~Glazman, and H.~U.~Baranger,
Phys.~Rev.~B {\bf 54}, 5637 (1996).

\bibitem{levitov} D.~A.~Abanin and L.~S.~Levitov,
Phys.~Rev.~Lett. {\bf 93}, 126802 (2004);
Phys.~Rev.~Lett. {\bf 94}, 186803 (2005).

\bibitem{dieterweiss} D.~Weiss, W.~Wegscheider, and U.~R\"o{\ss}ler, private communication.

\end{thebibliography}
\end{document}